\documentclass[%
 superscriptaddress,
 reprint,
 showpacs,preprintnumbers,
 nofootinbib,
 amsmath,amssymb,
 aps,
 prd,
 floatfix,
]{revtex4-1}
\usepackage{graphicx}% Include figure files
\usepackage{subcaption}
\usepackage{tikz}
\usepackage{xcolor}
\usepackage{hyperref}
\usepackage{comment}
\hypersetup{
    colorlinks=true,
    linkcolor=blue,
    filecolor=blue,      
    urlcolor=blue,
    pdftitle={FSI with INCL and NuWro},
    }

\usepackage{amssymb,amsmath,amstext,amsthm,amsfonts}
\usepackage[utf8]{inputenc}
\usepackage{float}
\usepackage{url}
\usepackage{multirow}
\usepackage{soul}
\newcommand{\pt}{p_\textrm{T}}

\newcommand{\dalphat}{\delta\alpha_\textrm{T}}

\newcommand{\dpt}{\delta \pt}

\RequirePackage{xspace}

\usepackage{lineno}
\definecolor{LightCyan}{rgb}{0.88,1,1}
\usepackage{xcolor,colortbl}
\usetikzlibrary{calc,patterns,angles,quotes}

\begin{document}

	\title{The role of de-excitation in the final-state interactions of protons in neutrino-nucleus interactions}

%%%%%%%%%%%%%%%%%%%%%%%%%%%%%%%%%%%%%%%%%%%%%%%%%%%%%%%%%%%%%%%%%%%%%%%%%%%%%%
    \author{A.~Ershova}
	\email[Contact e-mail: ]{anna.ershova@cea.fr}
	\affiliation{IRFU, CEA, Universit\'e Paris-Saclay, Gif-sur-Yvette, France}

 	\author{K. Niewczas}
	\affiliation{University of Wrocław, Institute of Theoretical Physics, Plac Maxa Borna 9, 50-204 Wrocław, Poland}
	\affiliation{Department of Physics and Astronomy, Ghent University, Proeftuinstraat 86, B-9000 Gent, Belgium}
		
	\author{S.~Bolognesi}
	\email[Contact e-mail: ]{sara.bolognesi@cea.fr}
	\affiliation{IRFU, CEA, Universit\'e Paris-Saclay, Gif-sur-Yvette, France}
	
	\author{A.~Letourneau}
	\email[Contact e-mail: ]{alain.letourneau@cea.fr}
	\affiliation{IRFU, CEA, Universit\'e Paris-Saclay, Gif-sur-Yvette, France}
	
	\author{J.-C. David}
	\affiliation{IRFU, CEA, Universit\'e Paris-Saclay, Gif-sur-Yvette, France}

    \author{J.~L.~Rodr\'iguez-S\'anchez}
	\affiliation{CITENI, Campus Industrial de Ferrol, Universidade da Coru$\tilde{n}$a, E-15403 Ferrol, Spain}
	
	\author{J.T. Sobczyk}
	\affiliation{University of Wrocław, Institute of Theoretical Physics, Plac Maxa Borna 9, 50-204 Wrocław, Poland}
	
	\author{A.~Blanchet}
	\affiliation{University of Geneva, Section de Physique, DPNC, Geneva, Switzerland}

    \author{M.~Buizza Avanzini}
	\affiliation{Laboratoire Leprince-Ringuet, CNRS, Ecole polytechnique, Institut Polytechnique de Paris, Palaiseau, France}
	
    \author{J.~Chakrani}
	\affiliation{Laboratoire Leprince-Ringuet, CNRS, Ecole polytechnique, Institut Polytechnique de Paris, Palaiseau, France}

    \author{J.~Cugnon}
	\affiliation{AGO department, University of Liège, all\'ee du 6 août 19, bâtiment B5, B-4000 Liège, Belgium}

 	\author{S.~Dolan}
	\affiliation{European Organization for Nuclear Research (CERN), 1211 Geneva 23, Switzerland}

	\author{C.~Giganti}
	\affiliation{LPNHE, Sorbonne Universit\'e, CNRS/IN2P3, Paris, France}

	\author{S.~Hassani}
	\affiliation{IRFU, CEA, Universit\'e Paris-Saclay, Gif-sur-Yvette, France}

    \author{J.~Hirtz}    
	\affiliation{IRFU, CEA, Universit\'e Paris-Saclay, Gif-sur-Yvette, France}

    \author{S.~Joshi}
	\affiliation{IRFU, CEA, Universit\'e Paris-Saclay, Gif-sur-Yvette, France}

   \author{C.~Juszczak}
	\affiliation{University of Wrocław, Institute of Theoretical Physics, Plac Maxa Borna 9, 50-204 Wrocław, Poland}

   \author{L.~Munteanu }
	\affiliation{European Organization for Nuclear Research (CERN), 1211 Geneva 23, Switzerland}
 
  \author{D.~Sgalaberna }
	\affiliation{ETH Zurich, Institute for Particle Physics and Astrophysics, Zurich, Switzerland}

\author{ U.~Yevarouskaya}
	\affiliation{LPNHE, Sorbonne Universit\'e, CNRS/IN2P3, Paris, France}

\begin{abstract}
\noindent 
Present and next generation of long-baseline accelerator experiments are bringing the measurement of neutrino oscillations into the precision era with ever-increasing statistics.  One of the most challenging aspects of achieving such measurements is developing relevant systematic uncertainties in the modeling of nuclear effects in neutrino-nucleus interactions. To address this problem, state-of-the-art detectors are being developed to extract detailed information about all particles produced in neutrino interactions. To fully profit from these experimental advancements, it is essential to have reliable models of propagation of the outgoing hadrons through nuclear matter able to predict how the energy is distributed between all the final-state observed particles. In this article, we investigate the role of nuclear de-excitation in neutrino-nucleus scattering using two Monte Carlo cascade models: NuWro and INCL coupled with the de-excitation code ABLA. The ablation model ABLA is used here for the first time to model de-excitation in neutrino interactions. As input to ABLA, we develop a consistent simulation of nuclear excitation energy tuned to electron-scattering data.
The paper includes the characterization of the leading proton kinematics and of the nuclear cluster production during cascade and de-excitation. The observability of nuclear clusters as vertex activity and their role in a precise neutrino energy reconstruction is quantified.

\end{abstract}

\maketitle

\section{Introduction}
\label{sec:introduction}

Neutrino oscillations have been discovered by measuring atmospheric and solar neutrinos and confronting them with respective flux predictions. Since then, the model of neutrino oscillations based on the PMNS mixing matrix has been refined thanks to measurements of artificially produced neutrinos with reactors and dedicated accelerators, as well as ever-increasing statistics of atmospheric neutrinos. In particular, long-baseline accelerator experiments are entrenched in the combined measurement of neutrinos before and after oscillations with so-called near and far detectors. The present-generation experiments (T2K~\cite{T2K:2023smv} and NOVA~\cite{Acero_2022}) are bringing the neutrino oscillation paradigm into the precision era while also addressing some still unknown parameters: the degree of Charge-Parity violation in neutrino oscillation, the ordering of neutrino masses (called normal, if mirroring the charged lepton mass ordering, or inverted otherwise), and the octant of the $\theta_{23}$ mixing angle~\cite{Cabrera:2020own, PhysRevD.103.112010}.
The next generation of long-baseline accelerator experiments (DUNE~\cite{Abi_2020} and Hyper-Kamiokande~\cite{Abe:2018uyc}) have the potential of definitive, high-statistics measurements of those unknown parameters. The success of such a program strongly depends on the capability of improving the control of systematic errors in neutrino oscillation measurements, notably those related to nuclear effects in neutrino-nucleus interactions. Such systematic uncertainties affect the kinematics of the final state particles, which serve as a proxy to reconstruct the neutrino energy and our ability to compare near and far detector data to extract neutrino oscillation measurements. 

To address the challenge of improved precision, the long-baseline experiments are moving from inclusive analyses, focused on the leptonic part of the neutrino-nucleus interaction final state, to exclusive analyses, including the hadronic component of the final state. To this aim, relatively new technologies for the field are being deployed, like using Liquid-Argon Time Projection Chambers in the SBN program~\cite{Antonello:2015lea} or the highly granular scintillator detector as the target in the upgraded T2K near detector~\cite{T2K:2019bbb}. The aim is to 
exploit detailed information on the hadronic final state to improve the understanding of nuclear effects: notably, in the quasi-elastic (QE) channel, the measurement of the final-state proton(s) could bring vital information. Whilst along this effort, a lot of attention has been devoted to the primary neutrino-nucleus interaction~\cite{NuSTEC:2017hzk}, but very few studies are available that highlight the impact of final-state interactions (FSI) on the outgoing particles in the nuclear matter, before leaving the nucleus. Advanced models, based on the mean-field picture of nuclear dynamics (e.g., Relativistic Mean-Field), are capable of a full quantum-mechanics description, including the effect of nuclear potential on the final state directly in the neutrino interaction modeling~\cite{Martinez:2005xe,Gonzalez-Jimenez:2019ejf}. Still, all available Monte Carlo simulations are based on a two-step simulation, where the FSI are simulated with a semi-classical cascade mechanism following the neutrino interaction. Different Monte Carlo generators tend to implement similar cascade models, which makes it challenging to study and quantify the uncertainties in the FSI mechanism. Even further, we are unaware of any study on the role of nuclear de-excitation in shaping the hadronic final state of neutrino-nucleus interactions. 

We started to investigate the impact of FSI on the hadronic part of the quasi-elastic neutrino-Carbon interaction in our previous reference~\cite{Ershova:2022jah} by comparing NuWro~\cite{NuWroREPO} and IntraNuclear Cascade Li\`ege (INCL)~\cite{Boudard:2012wc} models. INCL offers an entirely different nuclear model, unlike the other cascade mechanisms implemented in neutrino interaction event generators. Indeed, INCL has been originally developed to describe the interactions of baryons, mesons, and light nuclei on various target nuclei. Consequently, INCL also offers the compelling advantage of being systematically benchmarked to a large amount of hadron-nucleus scattering data~\cite{IAEA}.
In our previous study~\cite{Ershova:2022jah}, we highlighted essential differences between NuWro and INCL cascade models, and we characterized for the first time the production of nuclear clusters ($\alpha$, deutron, tritium, ...) in neutrino-nucleus interactions. In the present study, we push further the analysis by coupling INCL with the de-excitation code ABLA, thus simulating and characterizing the role of nuclear de-excitation in neutrino-nucleus QE interactions.

\section{Nuclear models}
\label{sec:models}

\subsection{NuWro}
\label{sec:nuwro}

NuWro is a versatile Monte Carlo event generator designed to study neutrino and electron interactions on nuclear targets for projectile energies ranging from $\sim$100~$\mathrm{MeV}$ to $\sim$100~$\mathrm{GeV}$~\cite{NuWroREPO}. In the case of scattering on nuclei, where applicable, simulations adopt the \textit{plane-wave impulse approximation} (PWIA) picture, making every interaction a two-step process: a primary interaction on bound nucleons, followed by hadron rescatterings (FSI). NuWro provides several dynamical mechanisms for the primary vertex, from the elastic or quasi-elastic reactions~\cite{Juszczak:2005wk}, through hyperon~\cite{Thorpe:2020tym} and single-pion production to deep-inelastic scattering~\cite{Juszczak:2005zs}. Additional channels such as two-body processes~\cite{Bonus:2020yrd}, coherent pion production~\cite{Berger:2008xs}, and neutrino scattering off atomic electrons~\cite{Zhuridov:2020hqu} are included for complex nuclear targets. Then, pions, nucleons, and hyperons are subject to FSI modeled with a custom intranuclear cascade model, which has been developed and constantly improved for over 15 years now~\cite{Golan:2012wx, Niewczas:2019fro, Thorpe:2020tym}. In the context of this work, we emphasize the technical aspects of modeling quasi-elastic neutrino-nucleus scattering in the used NuWro version (21.09).  One can find more information on aspects shared with former software versions in Refs.~\cite{Niewczas:2019fro, Niewczas:2020fev, Ershova:2022jah}.

Based on the PWIA picture, the calculation of the quasi-elastic scattering process factorizes into evaluating the \textit{hole spectral function} (SF) and the cross section on a bound nucleon target~\cite{Frullani:1984nn}. The former, denoted as $S(E,\vec{p})$, provides a probability of removing a bound nucleon of momentum $\vec p$ from the target nucleus while leaving the remnant nucleus in the state of energy
\begin{equation}
    E_R^\ast = M_A - M + E,
\end{equation}
where $M$ and $M_A$ are the rest masses of the target nucleon and nucleus, respectively, and $E$ is the argument of the spectral function. As an input to the factorized cross section, we use realistic spectral function profiles provided by \textit{O. Benhar et al.}~\cite{BENHAR1994493, Ankowski:2007uy}. This framework has been extensively studied in the context of exclusive electron scattering experiments~\cite{E97-006:2005jlg}, where the simultaneous detection of the final-state electron and knocked-out proton allows for measuring \textit{missing energy} $E_m$ and \textit{missing momentum} $\vec{p}_m$. These variables represent the energy and momentum deficit relative to the elastic electron-nucleon scattering case and provide much information about the nuclear structure. Missing energy is defined as
\begin{equation}
    E_m = M_R^\ast + M - M_A ,
\end{equation}
with
\begin{equation}
    M_R^\ast = \sqrt{(E_k + M_A - E_{k^\prime} - E_{p^\prime})^2 - |\vec{p}_m|^{2}} ,
\end{equation}
while missing momentum as
\begin{equation}
    \vec{p}_{m}=\vec{p^\prime} - \vec{k} + \vec{k^\prime}.
\end{equation}
Here, we denote the four-momenta of the projectile lepton, and outgoing lepton and proton as ($E_{k},\vec{k}$), ($E_{k^\prime},\vec{k^\prime}$), and ($E_{p^\prime},\vec{p^\prime}$), respectively. Because
\begin{equation}
    E_R^\ast = M_R^\ast + T_R ,    
\end{equation}
where $M_R^\ast$ and $T_R$ are the mass and kinetic energy of the excited remnant nucleus respectively, we get another interpretation of the argument of the spectral function: $E = E_m + T_R$. For light nuclei, the recoil $T_R$ is usually non-greater than a few MeV; thus, $E \simeq E_m$ and their distributions exhibit the same characteristics.

However efficient, the PWIA approach has limitations in precisely describing neutrino- and electron-nucleus scattering. Without treating the outgoing nucleon as a solution to the nuclear potential, as done in the \textit{distorted-wave impulse approximation} (DWIA), it is impossible to consistently account for many subtle effects~\cite{Nikolakopoulos:2019qcr, Gonzalez-Jimenez:2019ejf}, including the interaction phase space. Among them, the most meaningful is Pauli blocking, which constrains the allowed quantum states that the knocked-out nucleons may occupy. In NuWro, we resolve the Pauli blocking issue in the SF model by applying, on an event-by-event basis, the restriction based on the \textit{local Fermi gas}, i.e., with Fermi momentum as a function of local density. The nuclear density profile dictates the distribution of points where primary interactions occur. Compared to inclusive electron scattering, these results require further suppression of magnitude and a shift to lower energy transfer values, especially for forward scattering and lower projectile energies. One can obtain such an effect by following the procedure by \textit{A. Ankowski et al.}~\cite{Ankowski:2014yfa}, where the inclusive electron scattering cross section is folded as
\begin{equation}
\label{eq:sf_fsi}
    \frac{\mathrm{d}\sigma^{FSI}}{\mathrm{d}\omega \mathrm{d}\Omega} = \int \mathrm{d}\omega^\prime f(\omega - \omega^\prime - U_V) \frac{\mathrm{d}\sigma^{PWIA}}{\mathrm{d}\omega^\prime \mathrm{d}\Omega},
\end{equation}
where $\omega$ is the energy transfer, $\Omega$ is the solid angle of the outgoing lepton, and $U_V$ is the real part of the optical potential $U = U_V + iU_W$. The folding function $f$ can be decomposed as
\begin{equation}
    f(\omega) = \delta(\omega)\sqrt{\mathcal{T}} + \sqrt{1 - \mathcal{T}} \left( \frac{1}{\pi} \frac{U_W}{U_W^2 + \omega^2} \right),
\end{equation}
showing how the FSI-like effect of Eq.~\ref{eq:sf_fsi} is driven by the nuclear transparency $\mathcal{T}$ and the imaginary part of the optical potential $U_W$. 
Within this picture, the real part of the optical potential dictates the shift of the differential cross section peak, and the imaginary part determines its quenching and the associated enhancements of the distribution's tails, while the total cross section remains unchanged. This solution, introduced in NuWro 17.09, has been, together with the LFG-based Pauli blocking, the recommended setup of the spectral function model in NuWro.

In this work, we use the spectral function model together with INCL. Therefore, we need to make choices on the preferred configuration of NuWro. We note that both phenomena discussed above, i.e., the Pauli blocking and cross section folding, lead to a considerable fraction of events with the leading final-state proton of momentum below the Fermi momentum. To avoid possible ambiguities while using this model in our INCL implementation and to make our results consistent with the framework introduced in our previous work~\cite{Ershova:2022jah}, we refrain from using the folding procedure of Eq.~\ref{eq:sf_fsi}. Moreover, we perform Pauli blocking in the spectral function according to the global Fermi gas condition, i.e., with a constant Fermi momentum. To address these issues in the future development of this framework, we emphasize a need for consistent DWIA-based neutrino-nucleus interaction calculations implemented in Monte Carlo event generators~\cite{Nikolakopoulos:2022qkq}.

\subsection{INCL}
\label{sec:incl}

INCL has been originally developed to simulate the reactions of baryons (n, p, $\Lambda$, $\Sigma$), mesons (pions and Kaons), or light nuclei on a target nucleus. It demonstrates an exceptional consistency with various experimental data (see, for example, Refs.~\cite{2015, IAEA}). The INCL cascade is commonly followed by a de-excitation model, such as ABLA~\cite{Rodriguez-Sanchez:2022Abla,ABLA}, SMM~\cite{Botvina:1994vj, Bondorf:1995ua} or GEMINI++~\cite{Mancusi:2010tg, Charity:1988zz}. In this study, we have coupled INCL to the de-excitation model ABLA since it proved its 
applicability to nuclear interactions of the light carbon nucleus~\cite{ABLAlow}. Since the neutrino is not yet a projectile option in INCL, we use the neutrino vertex simulation provided by NuWro and inject into the FSI cascade simulation of INCL, as in Ref.~\cite{Ershova:2022jah}. 

INCL is mainly a classical model with a few extra components to simulate quantum effects. Each nucleon in the nucleus is assigned position and momentum and moves in the Woods-Saxon, modified-harmonic-oscillator (MHO), or Gaussian potential well, depending on the target nucleus characteristics~\cite{Rodriguez-Sanchez:2017odk}. Spectator nucleons do not interact with themselves to prevent the spontaneous boiling of the Fermi sea. The maximal Fermi momentum determines the radius of a sphere where nuclear momenta are equally distributed. In a classical picture, position and momentum have a one-to-one correlation. Taking into account the quantum properties of the wave functions, INCL employs a Hartree-Fock-Bogoliubov formalism and makes this correlation less strict. As a consequence, there is a non-zero chance that the nucleon will move beyond the maximum radius. Further details can be found in Ref.~\cite{Rodriguez-Sanchez:2017odk}.

Inside the INCL cascade, particles can decay (e.g., a $\Delta$ resonance or $\omega$ meson), interact with the nuclear medium, or attempt to leave the nucleus and subsequently either be deflected inside the nuclear medium or be ejected. While leaving the nucleus, particles can clusterize with the neighbor nucleons and leave as a nuclear cluster~\cite{Mancusi:2014fba}.

INCL features two options for Pauli blocking: the strict Pauli blocking model, which forbids interaction if the projectile momentum is lower than the Fermi momentum, and the statistical model, which includes only nearby nucleons in the phase-space volume and acts on the calculated occupation probability. In this study, the strict Pauli blocking should be applied to the neutrino interaction, but since it is modeled with NuWro, we adapt its method to this interaction, and the statistical Pauli blocking is subsequently used for following proton interactions. Another condition, the Coherent Dynamical Pauli Principle (CDPP)~\cite{PhysRevC.66.044615}, is applied to avoid problems resulting from the possible creation of holes in the Fermi sea during the initialization. Indeed, if holes exist, the local statistical Pauli method may allow for cascade events that will lead to a negative excitation energy.

To ensure the proper kinematics of the outgoing hadrons, the recoil of the residual nucleus is also calculated~\cite{Boudard:2012wc}. The iterative procedure is evoked to scale down the momenta of the outgoing hadrons. The recoil energy of the residual nucleus is not large (in 80\% of events, recoil energy is less than 2~MeV), but since carbon is a relatively light nucleus, the corresponding momentum, and therefore impact on the outgoing hadrons' kinematics has to be considered.

\begin{figure}[t!]
\centering
\includegraphics[width=0.98\linewidth]{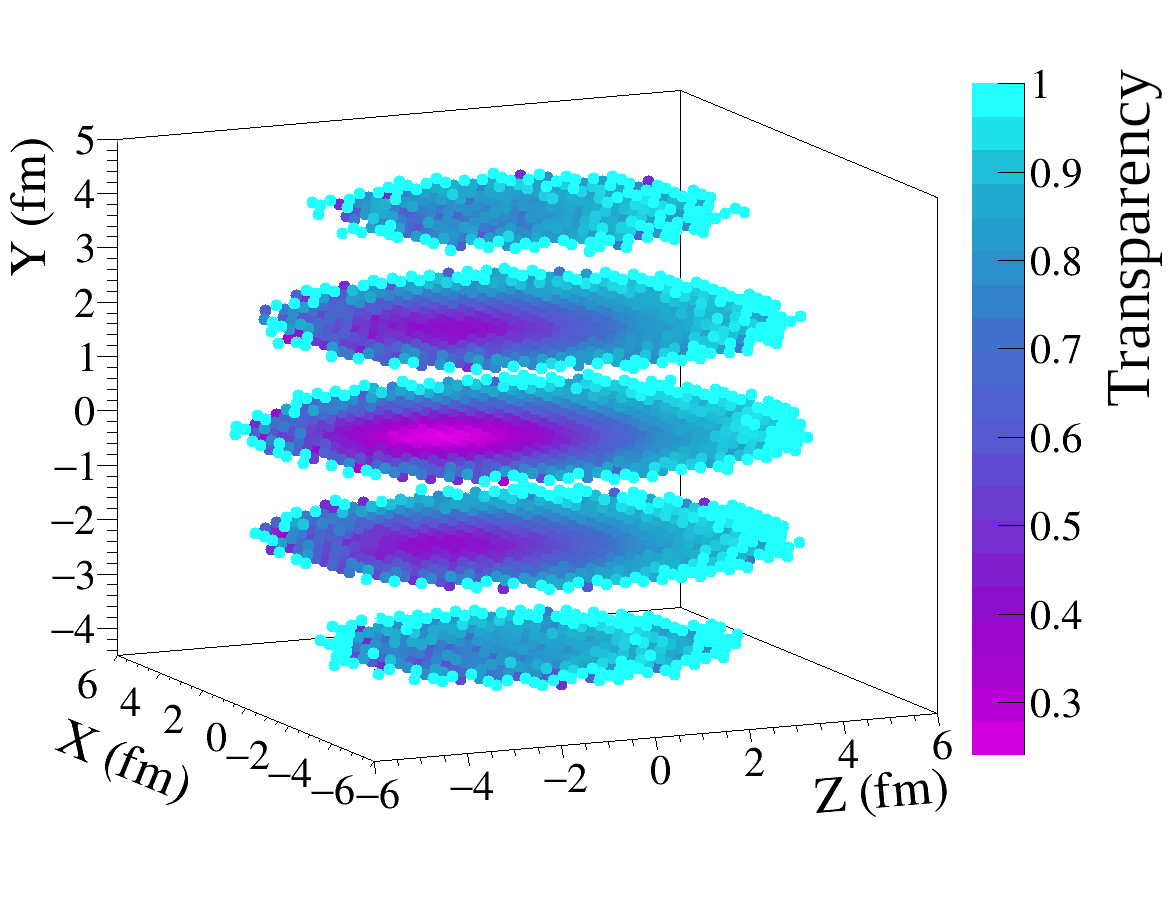}
\includegraphics[width=0.98\linewidth]{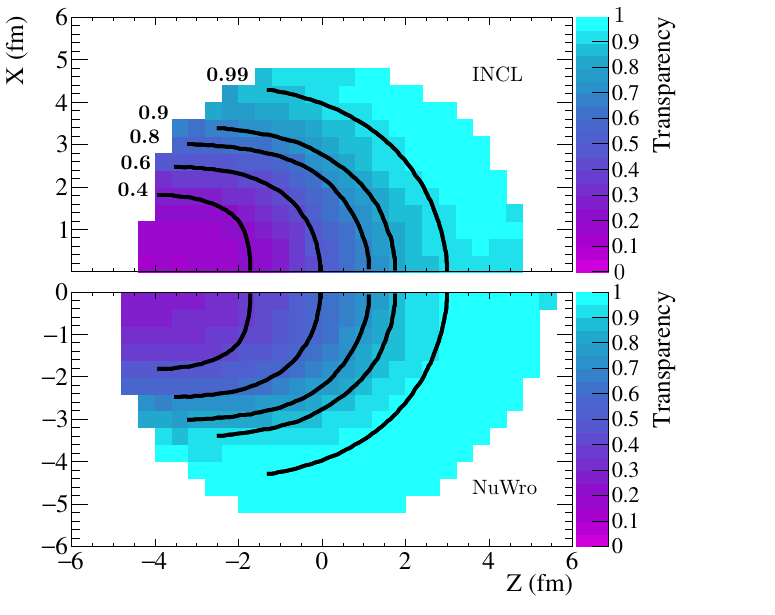}
\caption{\label{fig:transp} Top: nuclear transparency of $^{12}$C (ratio of a number of events without FSI to all events) depending on the position of the neutrino interaction inside the nucleus, simulated with INCL. The y-coordinate is averaged out to the 5 slices. The direction of the Z-axis corresponds to the neutrino direction. The center of the coordinate system is in the center of the nucleus. Bottom: nuclear transparency of $^{12}$C simulated with INCL and NuWro. The direction of the Z-axis corresponds to the outgoing proton direction. The center of the coordinate system is in the center of the nucleus. The solid lines are digitized from~\cite{Pandharipande:1992zz}.}
\end{figure}

In Fig.~\ref{fig:transp}, we present the nuclear transparency (the probability that the proton will leave the nucleus without re-interactions) depending on the position of the neutrino interaction. In the top panel, one can see that most of the transparent events originate downstream of the nucleus, where on average, nucleons propagate through the nuclear matter longer. Nuclear transparency is symmetric, with the Z-axis (neutrino direction) being an axis of symmetry. In the bottom panel, we compare transparency obtained with INCL and NuWro simulations with the lines of constant transparency from Ref.~\cite{Pandharipande:1992zz}. Here, the Z-axis corresponds to the proton direction. As expected, the less nuclear matter the proton passes through, the higher is transparency. As one can see, both INCL and NuWro quantitatively follow the same behavior as the theoretical model of Ref.~\cite{Pandharipande:1992zz}, with divergences due to different nuclear physics assumptions.

\subsection{ABLA}
\label{sec:abla}

The ablation model ABLA~\cite{Rodriguez-Sanchez:2022Abla} describes the de-excitation of an excited nuclear system through the emission of $\gamma$-rays, neutrons, light-charged particles, and intermediate-mass fragments (IMFs), or fission in case of hot and heavy remnants. The particle emission probabilities are calculated according to the Weisskopf-Ewing formalism~\cite{Weisskopf:1940}. Two phenomenological models, the constant temperature model of Gilbert-Cameron~\cite{Gilbert:1965} and the Fermi gas model based on the Bethe formula~\cite{Bethe:1937}, are used for the level-density calculations. Both approaches shift the excitation energy to consider the shell and pairing corrections~\cite{Moller:1995}. Additionally, to account for the role of collective excitations in the decay of excited remnants, the level density is corrected using vibrational and rotational enhancement factors~\cite{Junghans:1998}. Particle separation energies and emission barriers for charged particles are obtained according to the atomic mass evaluation AME2016~\cite{Huang:2017} and the phenomenological prescription given by \textit{W.~Qu et al.}~\cite{QU:2011}, respectively. 

The emission of $\gamma$-rays occurs in the last de-excitation stage of the evaporation cascade process. By assuming the power approximation for the radioactive strength function~\cite{Axel:1962} and the constant-temperature model~\cite{Gilbert:1965}, the statistical $\gamma$-emission rate is calculated according to Ref.~\cite{Ignatyuk:2001}. The effects of $\gamma$-ray decay are evident in the strength of the even-odd staggering of the final products, as shown in Ref.~\cite{Ricciardi:2004}. On the other hand, the discrete $\gamma$-ray emission from the lower-lying levels is omitted in ABLA since this requires specific nuclear structure databases~\cite{Capote:2009}. Hence, in the following study, the part of the $\gamma$-ray emission is missing.

\subsection{Excitation energy treatment}

Coupling ABLA to our simulations requires careful handling of the remnant nucleus excitation energy, a numerical input to the de-excitation routines. It is understood as the difference between masses of the excited and ground states of nuclear remnant:
\begin{equation}
    E_x = M_R^\ast - M_R .
\end{equation}
NuWro has yet to provide dedicated models predicting the fate of the residual nuclear system. Still, it provides sufficient information about the final-state particles to, by applying energy and momentum conservation, derive the properties of the remnant nucleus on an event-by-event basis. Within such a framework, the de-excitation, particularly neutron emission, was a topic of a recent study of the KamLAND collaboration~\cite{KamLAND:2022ptk}, where NuWro was used as the primary simulation tool. However, the model developed in that study has not entered the official distribution.

\begin{figure}[ht]
\centering
\includegraphics[width=0.98\linewidth]{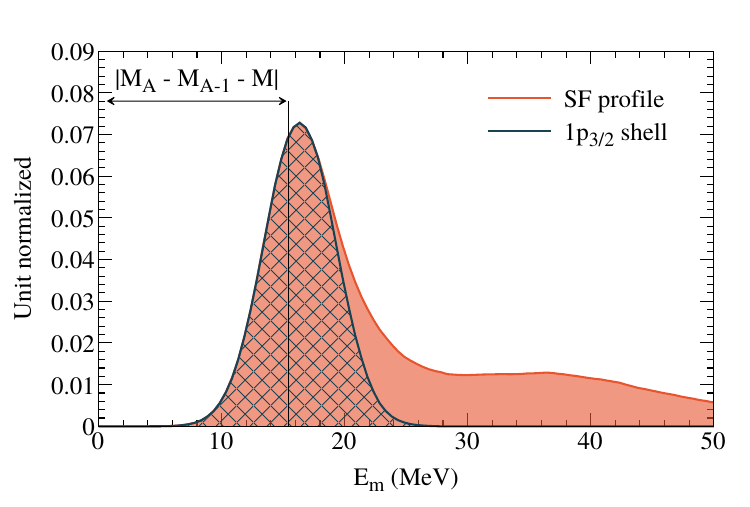}
\caption{\label{fig:Em} Missing energy profile extracted from the hole spectral function of Carbon~\cite{BENHAR1994493}. The area under the "SF profile" distribution is normalized to one. The highlighted area, labelled "$1p_{3/2}$", represents the expected contribution of valence nucleons, as described in the text.}
\end{figure}

While striving for consistency of our INCL implementation and the primary interaction model taken from NuWro, we need to ensure that the calculated value of nuclear excitation energy, which is coming from the primary neutrino interactions, reflects the properties of the target nucleon, as dictated by the used hole spectral function. Using the previously defined variables, we can write the excitation energy of one-nucleon knock-out as
\begin{equation}
    E_x = \sqrt{E_R^{\ast 2} - |\vec{P}_R|^2} - M_{A-1} ,
\end{equation}
where $M_{A-1}$ is the rest mass of the $A-1$ nucleus, and $\vec{P}_R = -\vec{p}_m$ is the momentum of the residual nucleus. The outcome of such a calculation depends strongly on the dynamics of target nucleons in our model. To obtain a more comprehensive interpretation, it is helpful to introduce the experimental definition of excitation energy:
\begin{equation}
    E_x^\mathrm{exp} = E_m - (M_A - M_{A-1} - M) .
\end{equation}
For carbon, the constant shift between the excitation and missing energies is $\sim 15.4 \ \mathrm{MeV}$. However, as presented in Fig.~\ref{fig:Em}, such a constant shift of missing energy leads to non-physical, negative values in our model. This effect of visible strength below the $1p_{3/2}$ peak value originates in the symmetric distribution used in the hole spectral function to describe the contribution of shells, using Saclay ($e,e^\prime p$) data as the basis for the spectral function construction~\cite{MOUGEY1976461}. 
To overcome this issue and properly evaluate the excitation energy for the $1p_{3/2}$ shell, we refer to high-precision measurements of the excitation energy coming from valence nucleons knock-out of Ref.~\cite{deWittHuberts_1990}, which provide relative contributions of discrete energy states of the remnant nucleus. We extract the fraction of the hole spectral function coming from the valence, $1p_{3/2}$ shell by assuming that it contains the whole strength below the peak value of the missing energy profile and that its distribution is symmetric. Fig.~\ref{fig:Em} presents the extracted strength of the $1p_{3/2}$ shell in red. Finally, we obtain the probability of interaction on the $1p_{3/2}$ shell by evaluating the ratio of the $1p_{3/2}$ shell and the complete missing energy profile. For these events, we adapt the first three discrete excitation energy states in the $^{12}\mathrm{C}(e,e^\prime p)$ process provided in Ref.~\cite{deWittHuberts_1990}, i.e., $3/2^-$ (ground state, $E_x = 0~\mathrm{MeV}$) corresponds to $79 \%$ of events, $1/2^-$ ($2.125 \ \mathrm{MeV}$) to $12 \%$, and $3/2^-$ ($5.02 \ \mathrm{MeV}$) to $9 \%$. 
We refer to excitation energies calculated with this procedure as $E_x^\mathrm{SF}$. Additionally, to apply it to charged-current neutrino interactions on target neutrons, we incorporate a constant Coulomb correction of $2.8 \ \mathrm{MeV}$.

Further, INCL handles excitation energy calculation for events where the leading nucleon experiences final-state interactions with its dedicated routine.
It is derived in a standard manner as the difference between the total binding of the initial ($B_A$) and remnant ($B_R$) nuclei plus the separation energies of all knocked-out particles. By convention, we define the total binding for light nuclei ($A<56$) as a negative value; therefore, $M_A = A \cdot M + B_A$. One can calculate the binding in terms of nucleon constituents as
\begin{equation}
    B_A = \sum_i^{A}(E_i - M - V_i) = \sum_i^{A}(T_i - V_i) ,
\end{equation}
where $E_i$, $T_i$, and $V_i$ represent the total energy, kinetic energy, and potential of bound nucleons, respectively. Thus, in a typical reaction emitting $N$ nucleons, we can evaluate the excitation energy as
\begin{equation}
    E_x^\mathrm{INCL} =
    B_{A-N} - B_A + N \cdot E_s,
\end{equation}
where we treat the nucleon separation energy $E_s$ as constant, with a value of $6.8 \ \mathrm{MeV}$. Finally, in our implementation, we substitute the target nucleon properties ($T_n$ and $V_n$) in the following way
\begin{equation}
\label{eq:excitation_sf_incl}
    E_x^\mathrm{SF+INCL} = E_x^\mathrm{INCL}
    - (T_n - V_n + E_s) + E_x^\mathrm{SF} .
\end{equation}
Therefore, we ensure that in the limit of no FSI, we retain the PWIA result ($E_x^{\mathrm{SF+INCL}} = E_x^\mathrm{SF}$).

\begin{figure}[ht]
\centering
\includegraphics[width=0.98\linewidth]{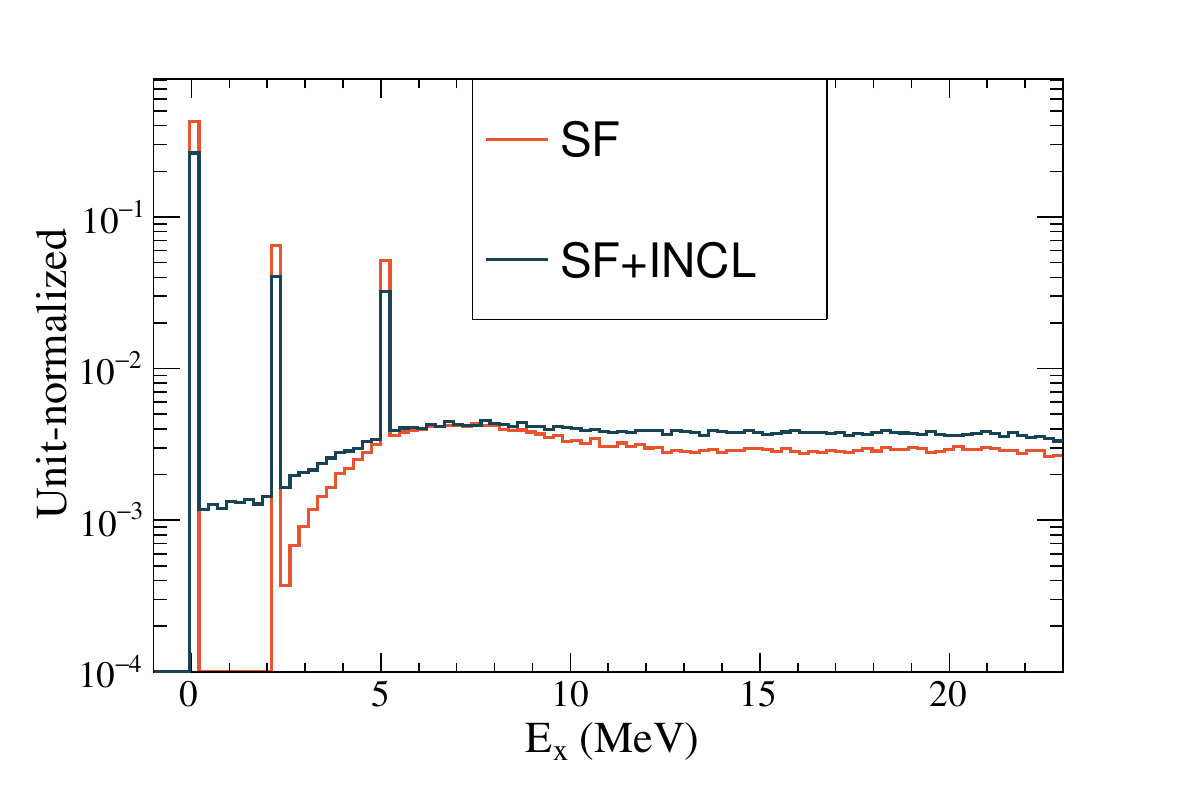}
\includegraphics[width=0.98\linewidth]{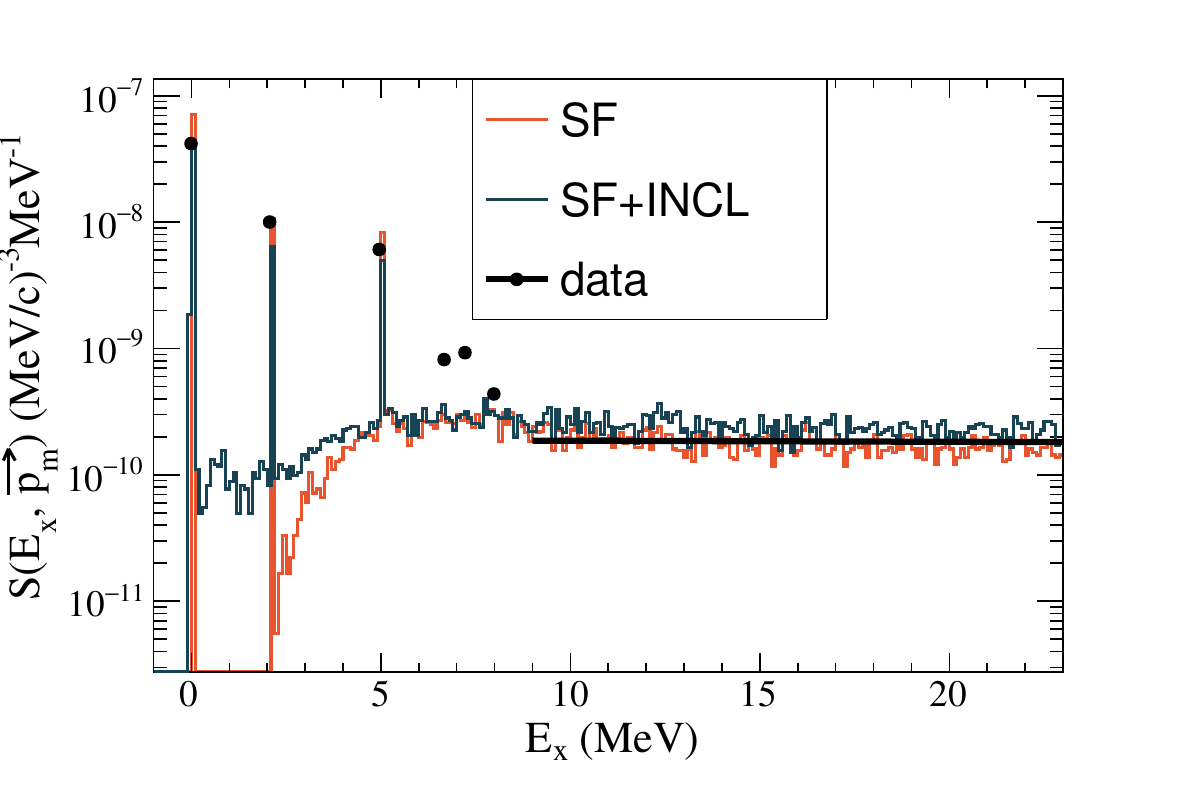}
\caption{\label{fig:Est} Top: excitation energy of the nuclear remnant after neutrino interactions. "SF" is the calculation in the pure PWIA approach, while "SF+INCL" is the result after the cascade, as described in the text. Bottom: excitation energy of the nuclear remnant after neutrino interactions for $169.5 < p_m < 174.5$~MeV, as presented in Ref.~\cite{VanDerSteenhoven:1988um}. Data are digitalized from Ref.~\cite{VanDerSteenhoven:1988um}. The s-shell contribution is approximated with a linear fit. The first three peaks were included in the model, while the next three were neglected.}
\end{figure}

We present the final results of our excitation energy calculations in Fig.~\ref{fig:Est}. One can see that the discrete states below $\sim 6 \ \mathrm{MeV}$ dominate the excitation landscape. Including final-state interactions redistributes this strength and flattens the $1s_{1/2}$-shell-dominated background. The bottom panel of Fig.~\ref{fig:Est} shows a remarkable accuracy while comparing these distributions to the experimental data of Ref.~\cite{VanDerSteenhoven:1988um}. For this comparison, we take the corresponding slices in terms of the average missing momentum. Moreover, it is worth mentioning that our methodology is similar to the algorithm applied in the aforementioned studies of the KamLAND collaboration~\cite{Abe:2021cgz}. However, we find our approach more exhaustive as it is consistent with the hole spectral function implemented in NuWro, with the addition of vital experimental input.

\section{Analysis and results}
\label{sec:an_res}

We focus on the Charged-Current Quasi-Elastic (CCQE) neutrino interactions on Carbon modeled with the T2K neutrino flux from Ref.~\cite{T2K:2015sqm}.

We simulate about 500,000 CCQE events with NuWro. Event by event, we inject into the INCL nuclear model the leading proton exiting the neutrino interaction and simulate the FSI cascade with INCL. The de-excitation simulation performed by ABLA follows the INCL cascade. Due to short-range correlations, there are two outgoing protons in the NuWro neutrino vertex in 15\% of events. In the INCL simulation, we keep only the leading proton that starts the cascade. We have tested that removing the SRC events does not affect the conclusions on the FSI characterization in NuWro and INCL.

We study the de-excitation impact on the leading proton (proton with the highest momentum in the event) kinematics with the Single Transverse Variabels (STV) on the nucleon multiplicity and on the nuclear cluster production.

\begin{figure*}[ht!]
\begin{minipage}[h]{0.48\linewidth}
\center{\includegraphics[width=1\linewidth]{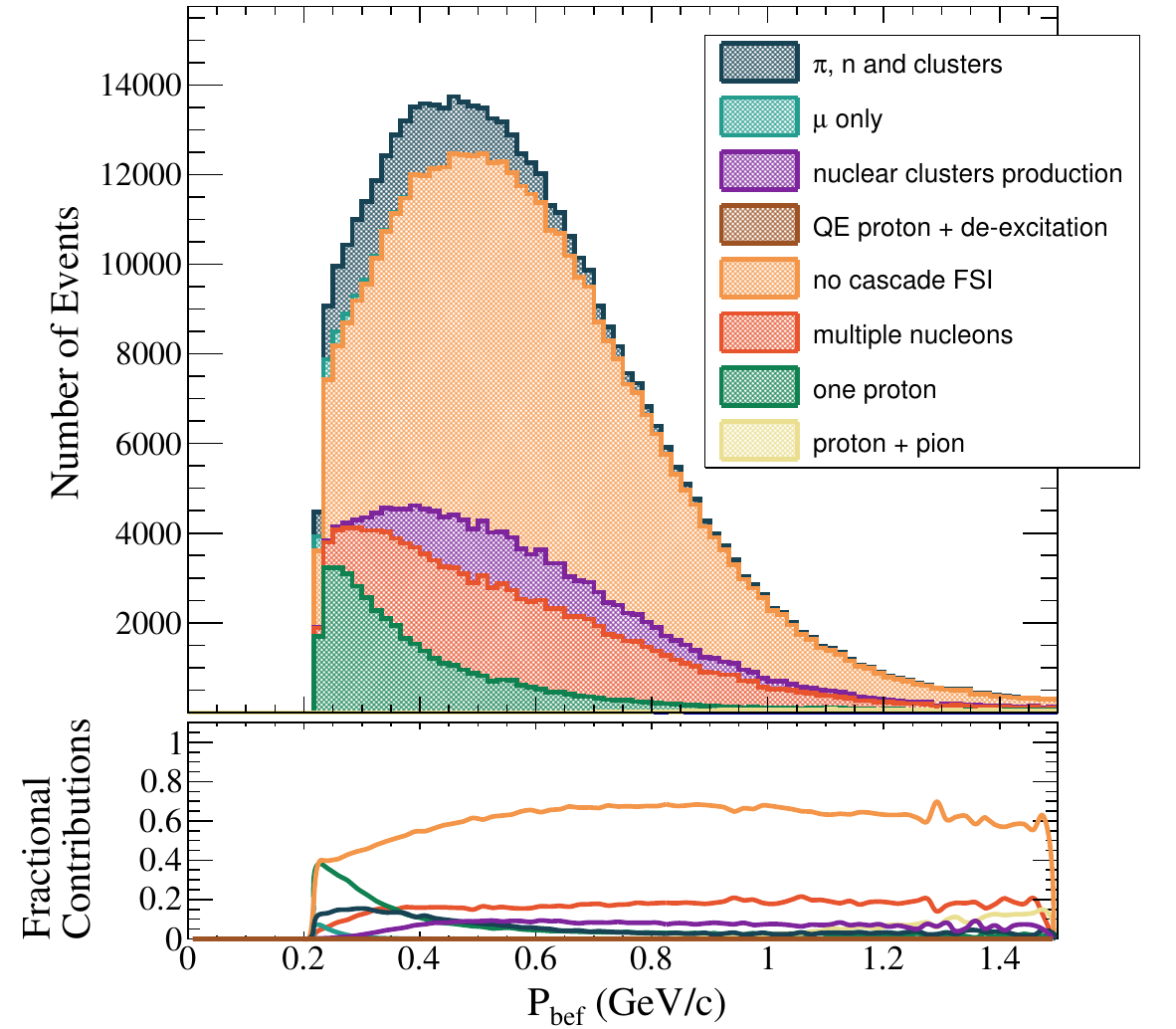} \\}
\end{minipage}
\hfill
\begin{minipage}[h]{0.48\linewidth}
\center{\includegraphics[width=1\linewidth]{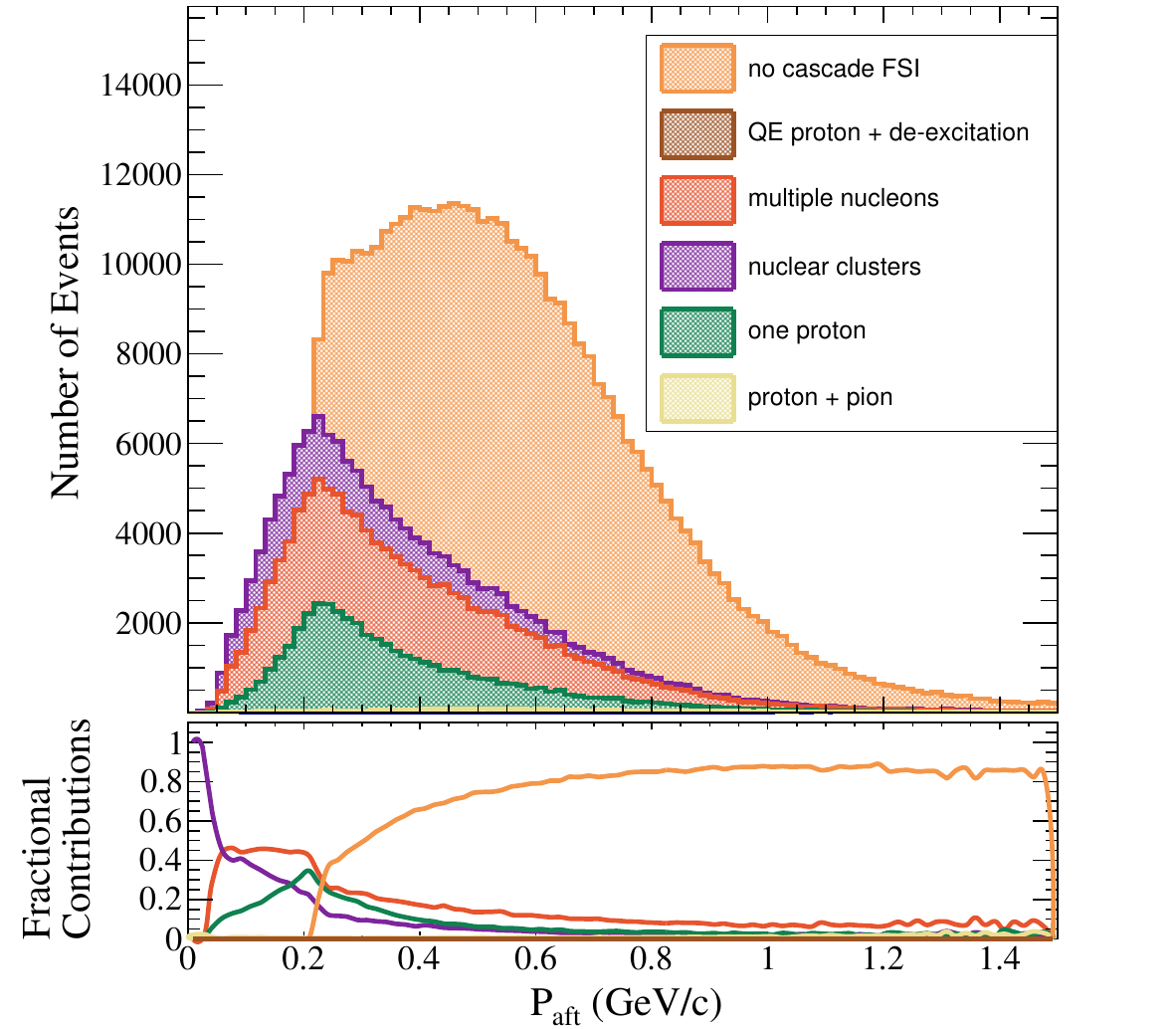} \\}
\end{minipage}
\vfill
\begin{minipage}[h]{0.48\linewidth}
\center{\includegraphics[width=1\linewidth]{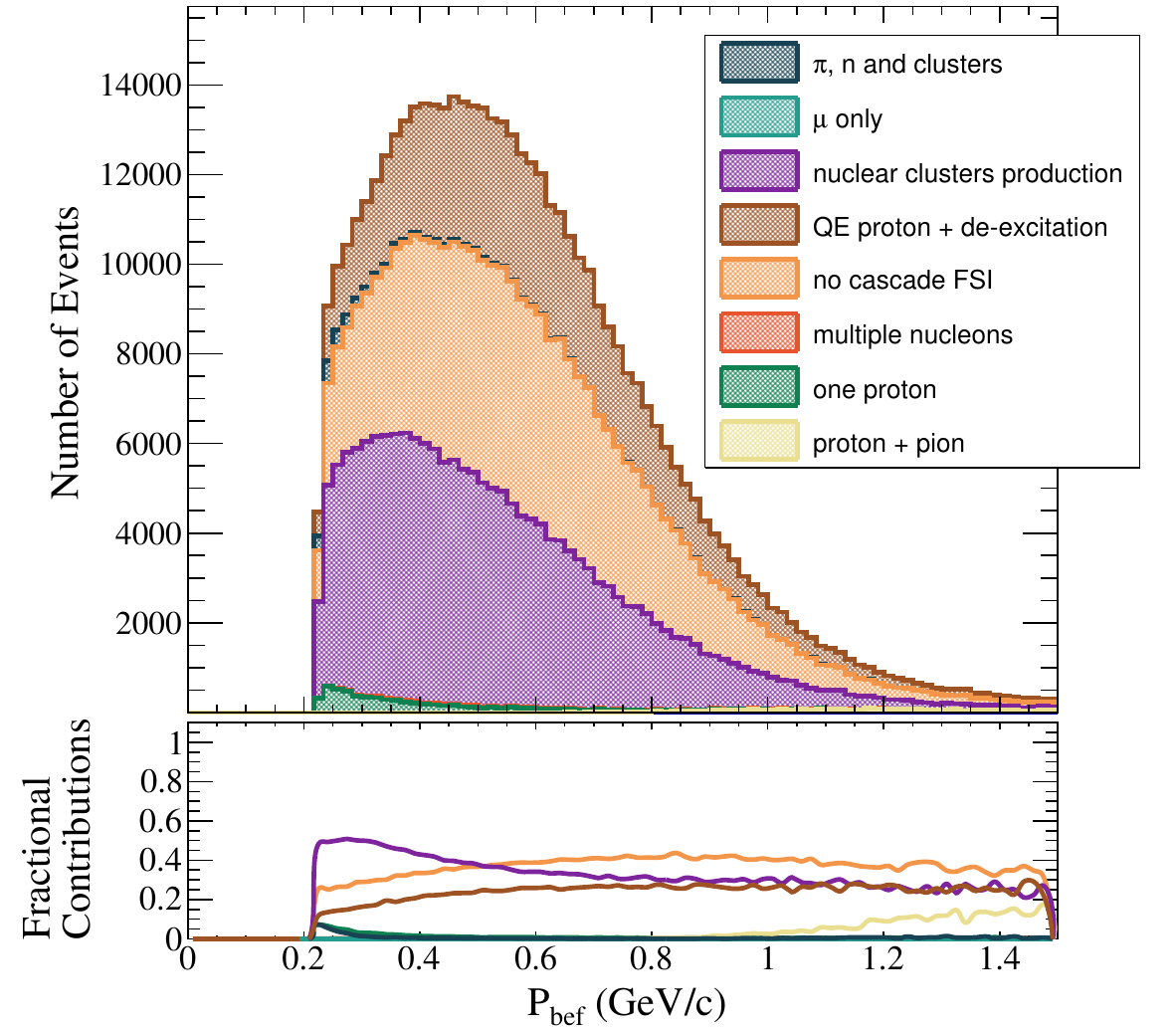} \\}
\end{minipage}
\hfill
\begin{minipage}[h]{0.48\linewidth}
\center{\includegraphics[width=1\linewidth]{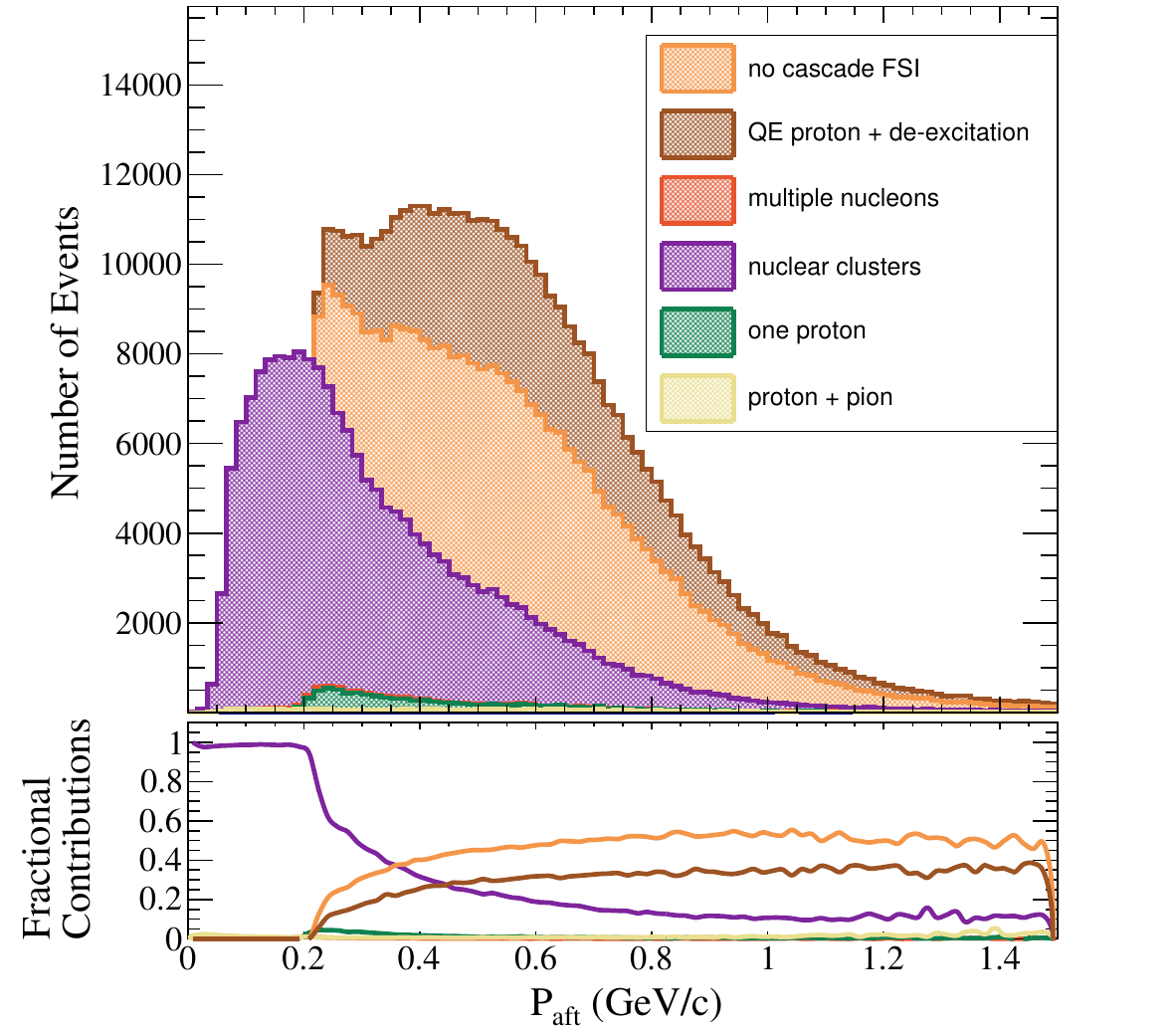} \\}
\end{minipage}
\vfill
\begin{minipage}[h]{0.48\linewidth}
\center{\includegraphics[width=1\linewidth]{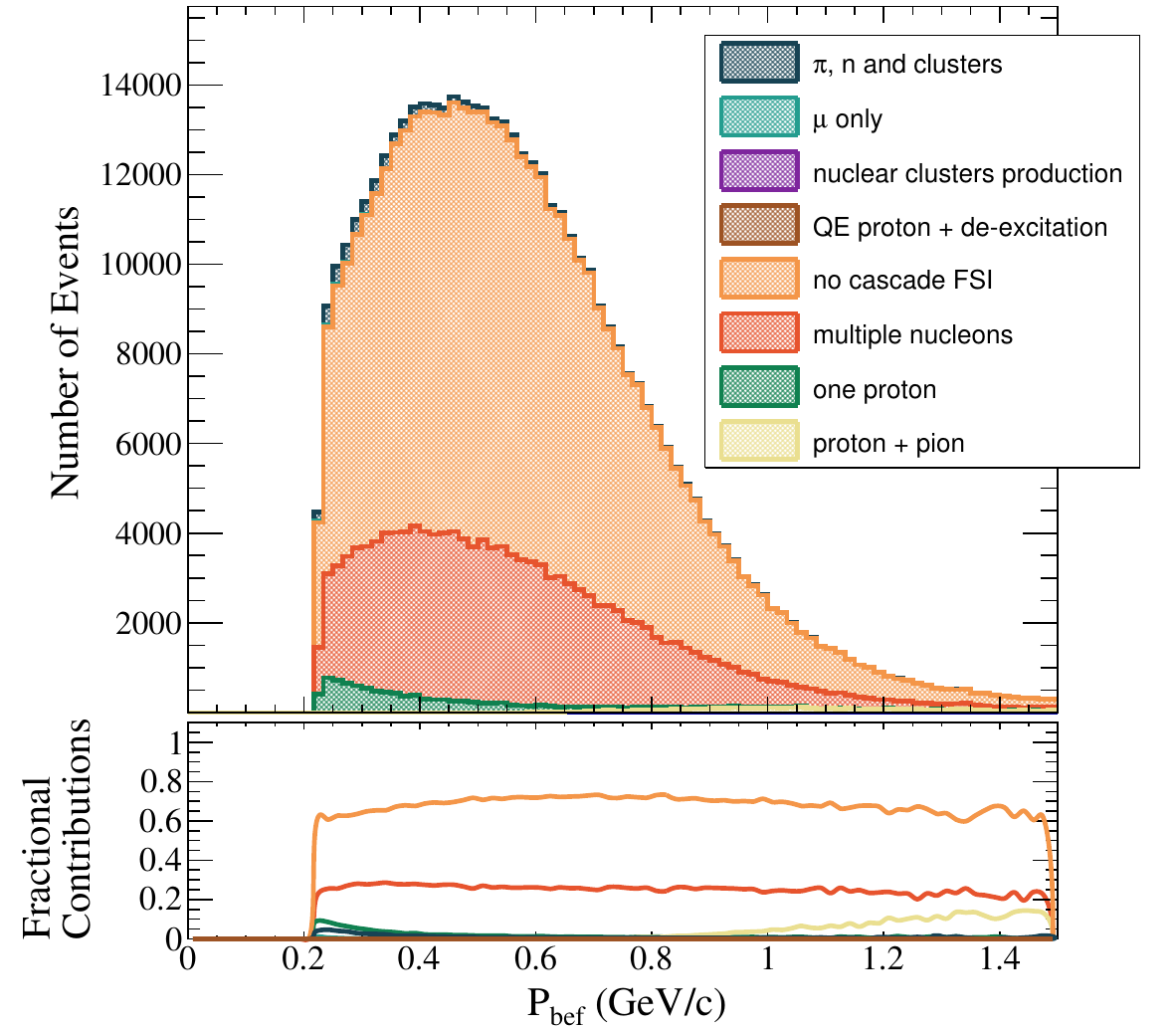} \\}
\end{minipage}
\hfill
\begin{minipage}[h]{0.48\linewidth}
\center{\includegraphics[width=1\linewidth]{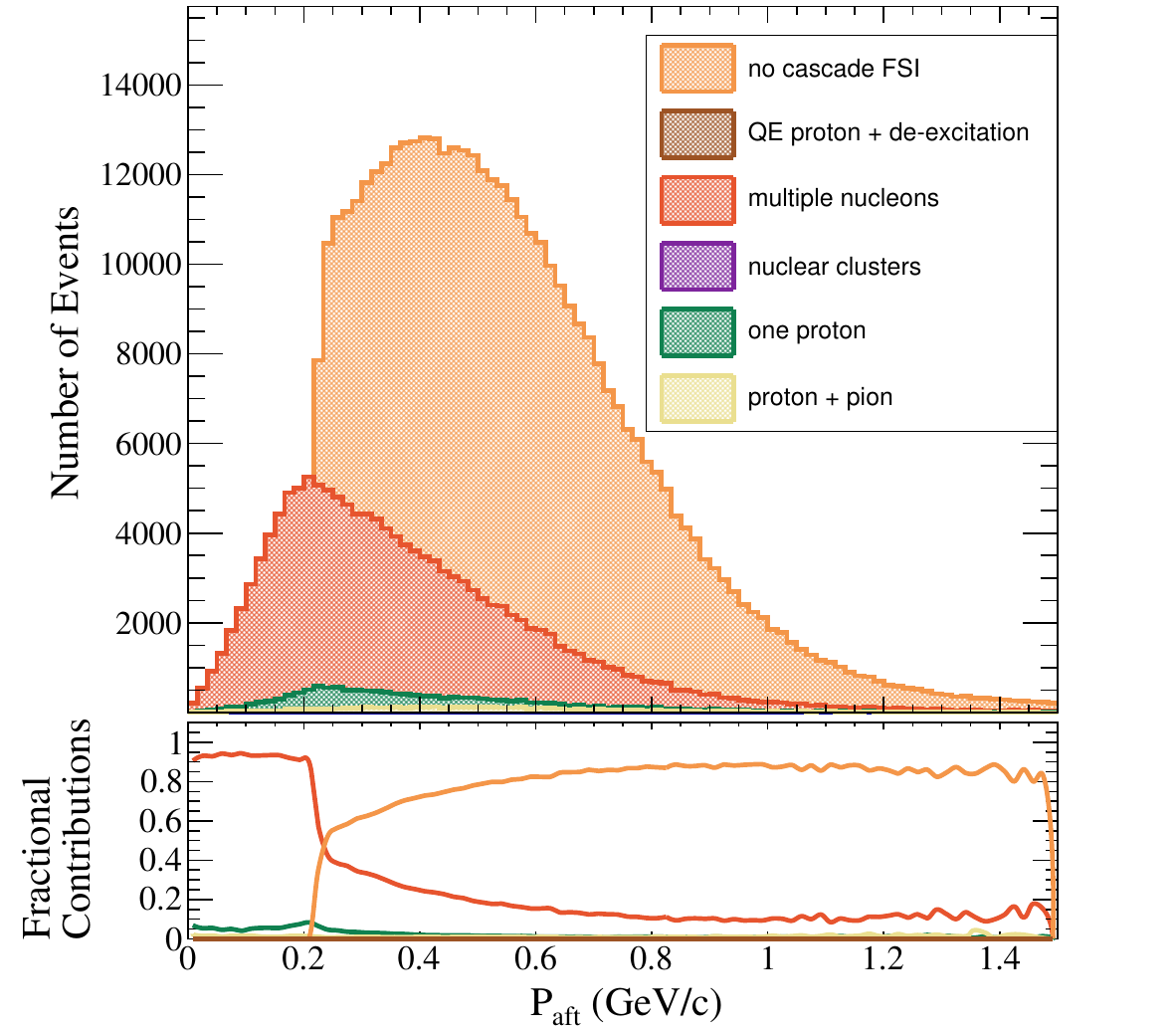} \\}
\end{minipage}
\caption{\label{fig:mom_bef} 
Proton momentum before (left) and after (right) FSI in CCQE events with T2K neutrino energy flux. Top: INCL, middle: INCL + ABLA, bottom: NuWro SF. The shape of proton
momentum before FSI is by definition identical for INCL and NuWro cascades. The 0 proton channel in NuWro includes muon only and pion and neutron production. There is no cluster production in NuWro. }
\end{figure*}

We present the leading proton momentum before and after FSI simulated with NuWro, INCL alone, and INCL+ABLA in Fig.~\ref{fig:mom_bef}. The shape of the proton momentum before FSI is identical for all models since it comes from the NuWro vertex simulation, and different colors display the fate of the leading proton. We distinguish a few channels with multiple particles emission based on the types of ejected particles. "No cascade FSI" includes events with no change of energy of the leading proton and no other particles produced during the cascade. We consider these events to be "transparent". "One proton" consists of events with only one proton in the final state with energy different from the proton energy before FSI. A "proton + pion" channel corresponds to a proton production and at least one pion in the final state. "$\pi$, n, and clusters" is a channel with events without protons but other particles being produced. "Multiple nucleons" contains events with various numbers of protons and neutrons produced. "Nuclear clusters production" consists of events in which at least one proton and a nuclear cluster leave the nucleus. "$\mu$ only" contains proton absorption events where only the muon left the nucleus. The last two channels are absent in the NuWro simulation. Results of the "$\mu$ only" channels have changed with respect to Ref.~\cite{Ershova:2022jah} since we updated the INCL treatment to better match the SF formalism implemented in NuWro. "QE proton + de-excitation" is a unique channel for the ABLA simulation: it corresponds to the situation where the leading proton left the nucleus without interaction, and other particles were produced during de-excitation. Here, excitation energy comes from the neutrino interaction. The bottom panels of plots in Fig.~\ref{fig:mom_bef} represent the relative fraction of each channel depending on the proton momentum. INCL+ABLA FSI channels are massively dominated by nuclear cluster production. Bare INCL cascade features a significant fraction of events with no proton in the final state, as was discussed in Ref.~\cite{Ershova:2022jah}. During de-excitation, more low-momentum particles (mainly protons) are produced, so INCL+ABLA simulation has a similar fraction of events with no proton in the final state as NuWro. Even though we have recovered some events with a proton in the final state, the kinematics of these protons are very different from NuWro since they were produced by the de-excitation and not by the FSI cascade. The "multiple nucleons" channel constitutes around~1\% of all events, while in NuWro, this channel corresponds to 26\% of events. A fraction of the "no cascade FSI" events remains in the INCL+ABLA simulation, stemming mostly from the events with the interaction on the $1p_{3/2}$ shell that results in zero excitation energy with 79\% chance, as can be seen in Fig.~\ref{fig:Est}. However, for the events that contribute to this channel, the excitation energy is not zero but lower than needed to remove nucleons, so gamma production should occur. Yet, as mentioned in section~\ref{sec:abla}, the discrete $\gamma$-emision is not handled by ABLA.

\begin{figure}[ht!]
\centering
\includegraphics[width=0.98\linewidth]{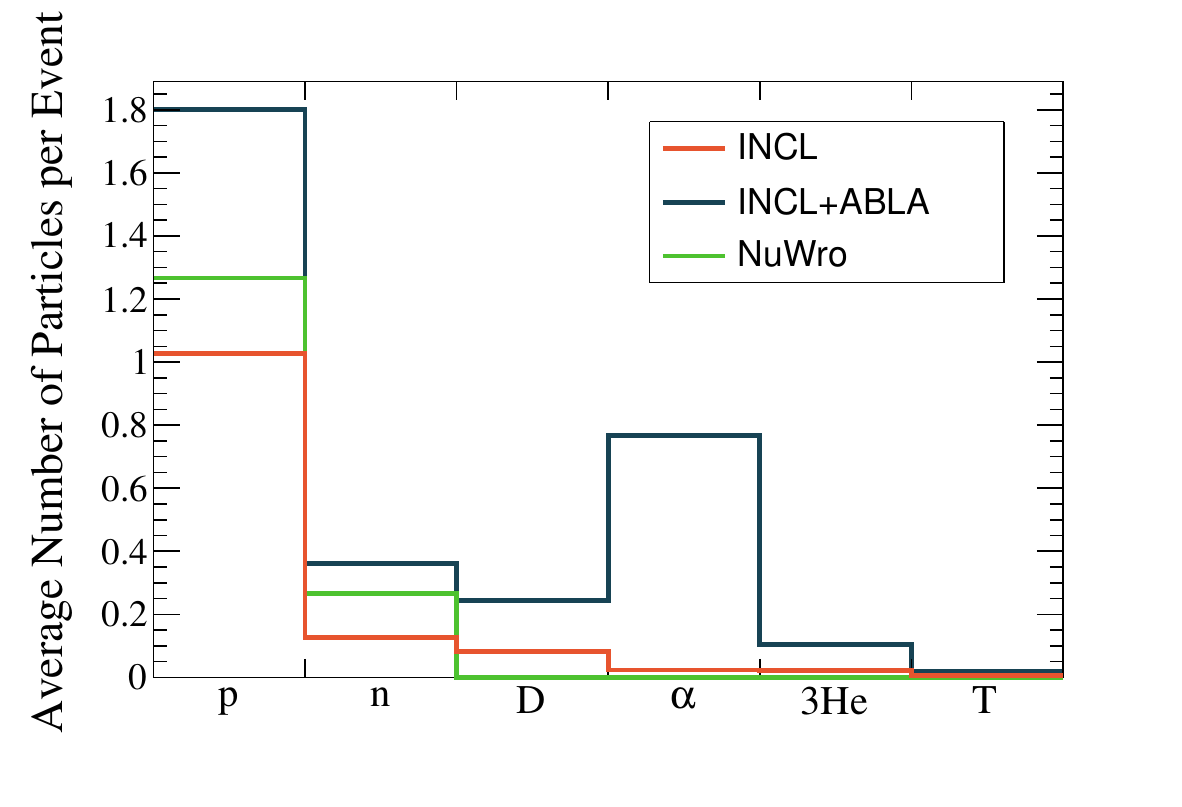}
\caption{\label{fig:part} Average number of particles produced per event for INCL, INCL coupled with ABLA and NuWro. NuWro produces only protons and neutrons.}
\end{figure}

Fig.~\ref{fig:part} shows the average number of particles per event produced by INCL, INCL+ABLA, and NuWro. NuWro produces more protons than INCL, but ABLA enhances proton production by factor 2. Also, ABLA increases the production of the $\alpha$ particles and neutrons by a few times.

We employ STV to characterize the leading proton kinematics after FSI. We will use the following STV~\cite{Lu:2015tcr}, which are used in the analysis of neutrino experiments:
\begin{equation}
\begin{gathered}
   \delta \alpha_{T} = arccos\frac{-\vec{p^{\mu}}_{T} \cdot \delta \vec{p}_{T}}{p^{\mu}_{T} \cdot \delta p_{T}} \\
   |\delta \vec{p_{T}}| = |\vec{p^{p}}_{T} + \vec{p^{\mu}}_{T}| \\
  % \delta \phi_{T} = arccos\frac{\vec{p^{\mu}}_{T} \cdot (\vec{p}_p)_{T}}{p^{\mu}_{T} \cdot (p_p)_{T}} ,
\end{gathered}
\end{equation}
where $\vec{p^{p}}_{T}$ is the component of the proton momentum projected into the plane transverse to the neutrino direction (transverse component) and $\vec{p^{\mu}}_{T}$ is the transverse component of the muon momentum. 

The variable $\dalphat$ (the transverse boosting angle) is particularly sensitive to the leading proton FSI. The $\dalphat$ distribution is supposed to be uniform for transparent events. In the case of FSI that generally decelerates the outgoing particles, we expect an enhancement of the $\dalphat$ distribution in the high $\dalphat$ region ($\dalphat$\textgreater90$^o$).

In the case of neutrino interaction on a nucleon at rest, $\dpt$ equals zero. For transparent events, it represents the Fermi motion distribution. FSI tends to induce further unbalancing between muon and proton momentum and thus increase $\dpt$ and might shift the peak of the distribution and contribute to the high energetic tail. 

\begin{figure*}[ht!]
\centering
\begin{minipage}[h]{0.48\linewidth}
\center{\includegraphics[width=1\linewidth]{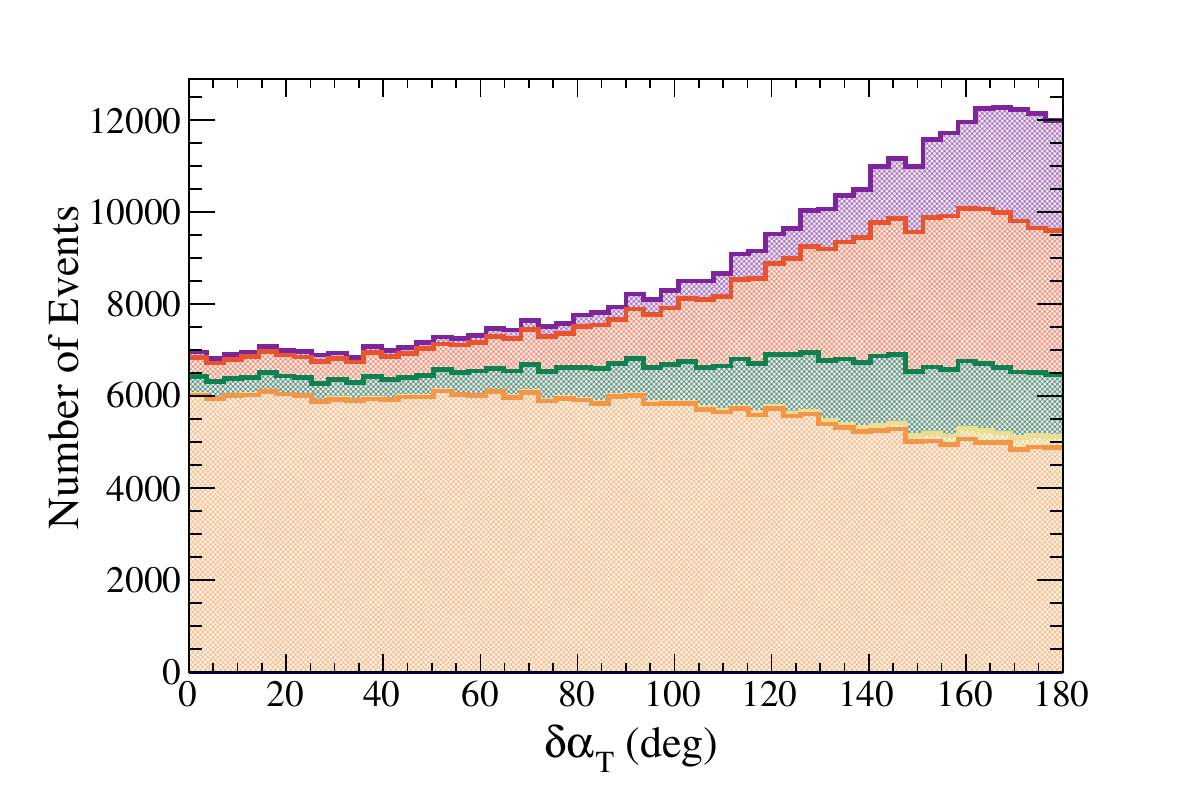} \\}
\end{minipage}
\hfill
\begin{minipage}[h]{0.48\linewidth}
\center{\includegraphics[width=1\linewidth]{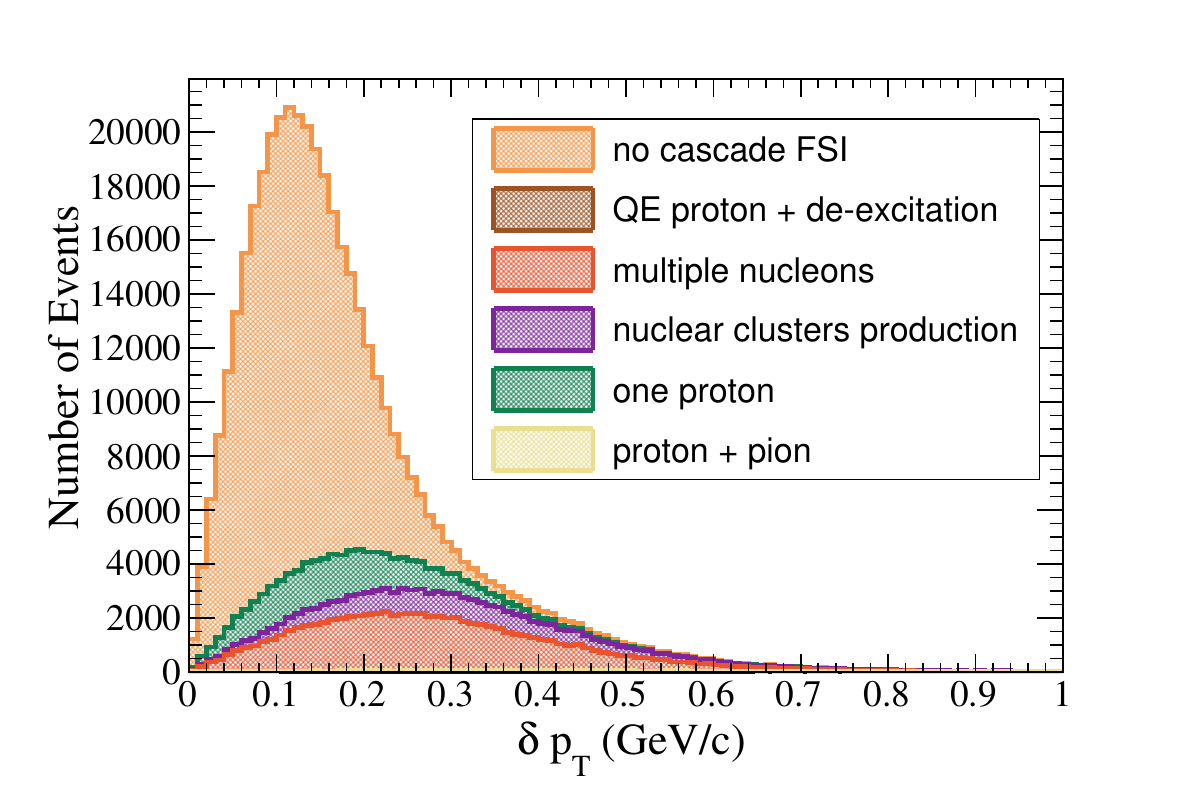} \\}
\end{minipage}
\vfill
\begin{minipage}[h]{0.48\linewidth}
\center{\includegraphics[width=1\linewidth]{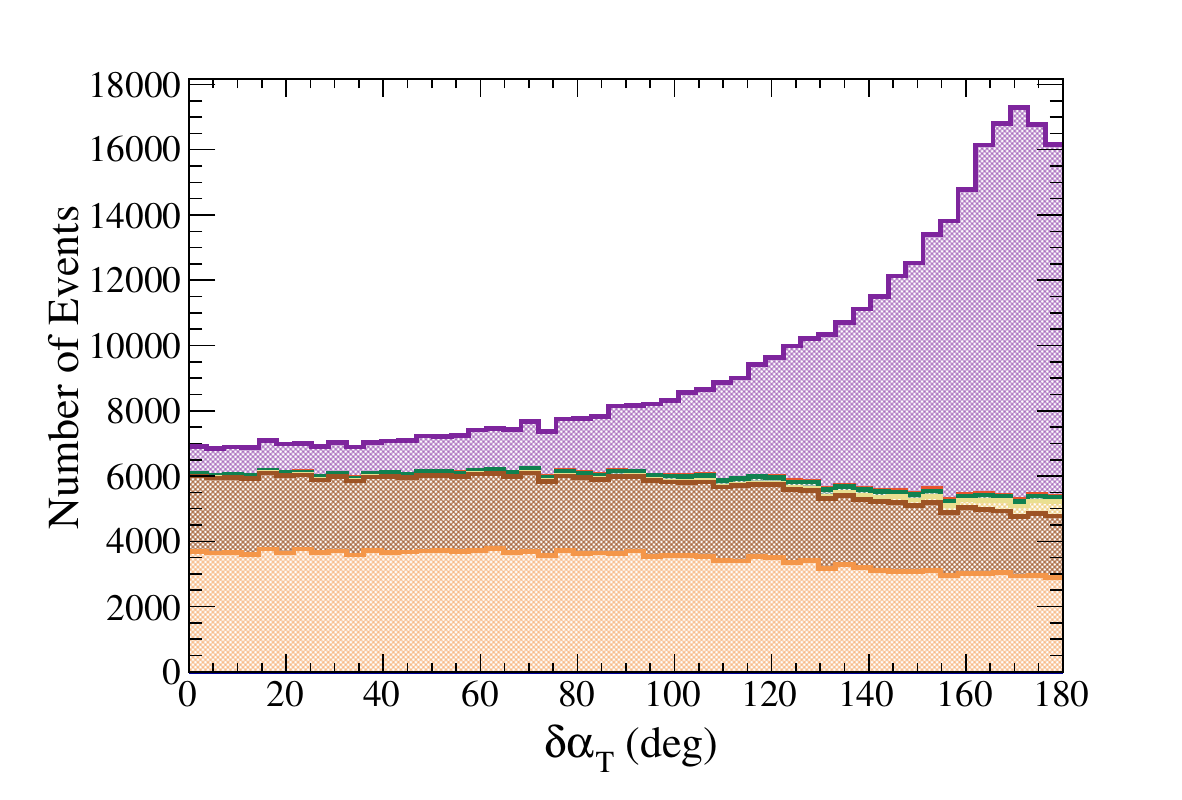} \\}
\end{minipage}
\hfill
\begin{minipage}[h]{0.48\linewidth}
\center{\includegraphics[width=1\linewidth]{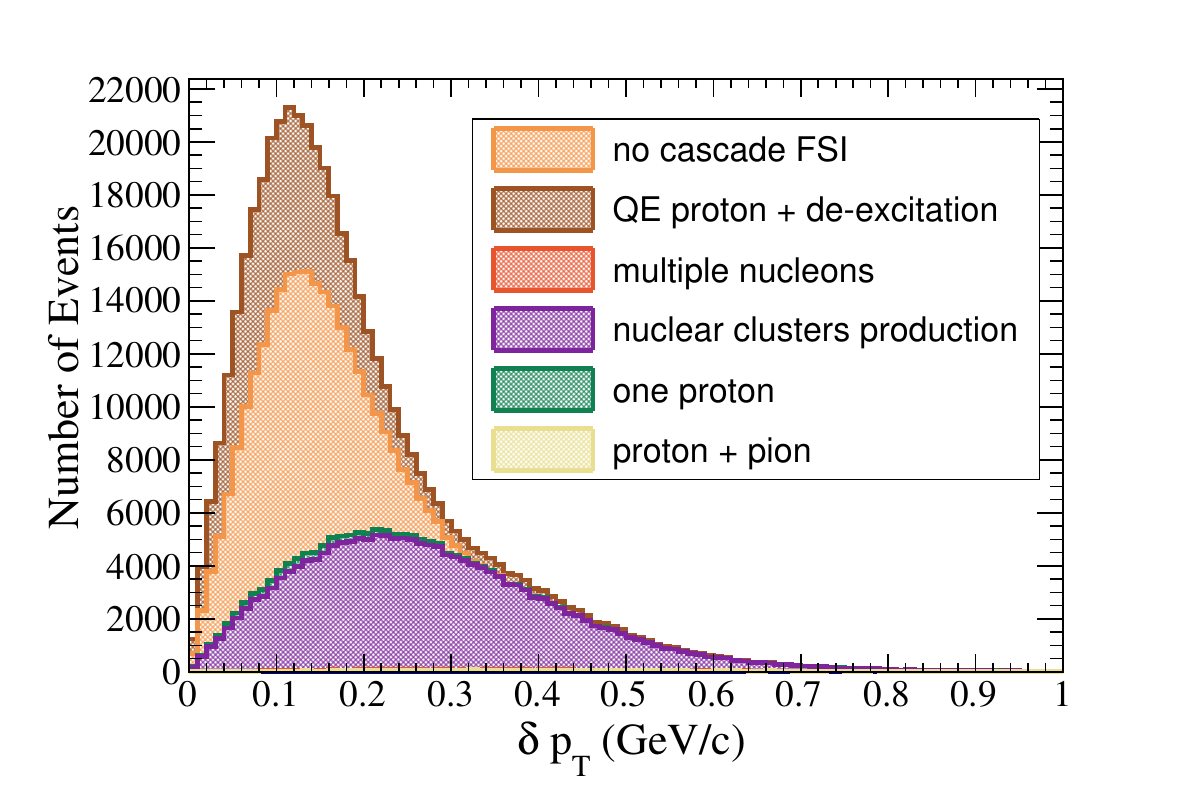} \\}
\end{minipage}
\vfill
\begin{minipage}[h]{0.48\linewidth}
\center{\includegraphics[width=1\linewidth]{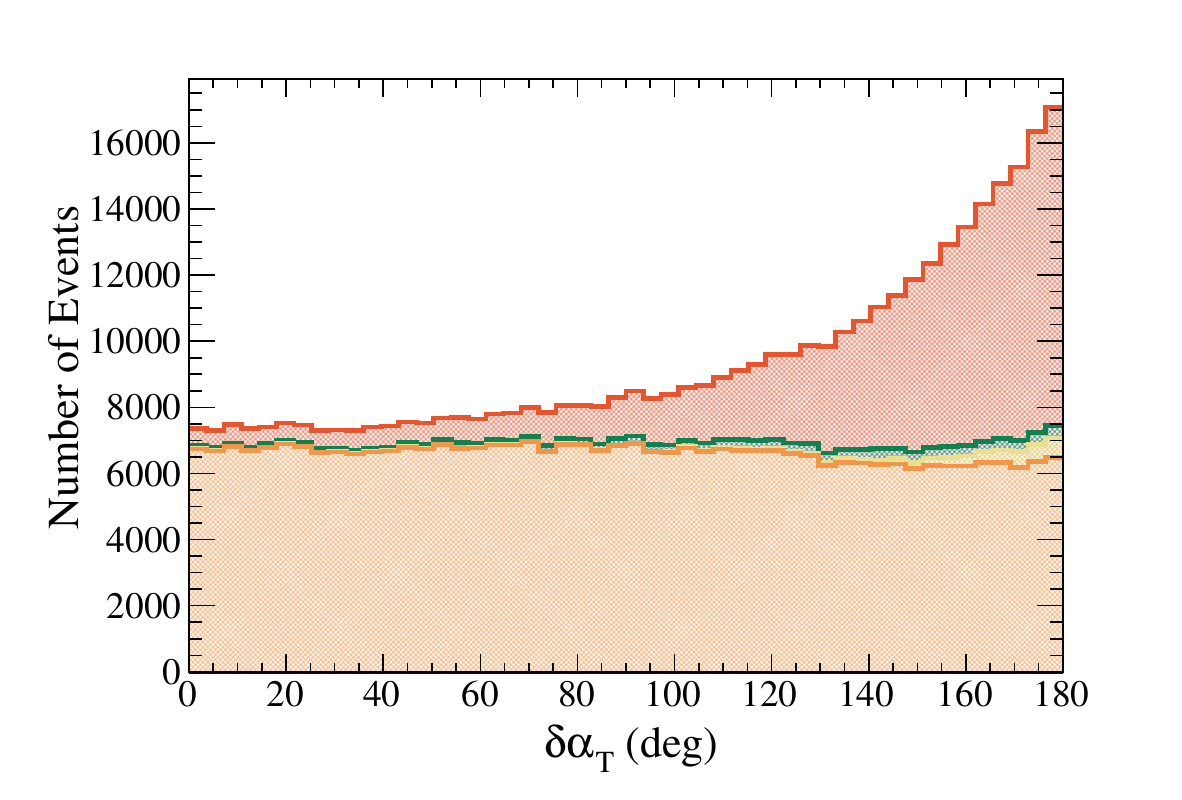} \\}
\end{minipage}
\hfill
\begin{minipage}[h]{0.48\linewidth}
\center{\includegraphics[width=1\linewidth]{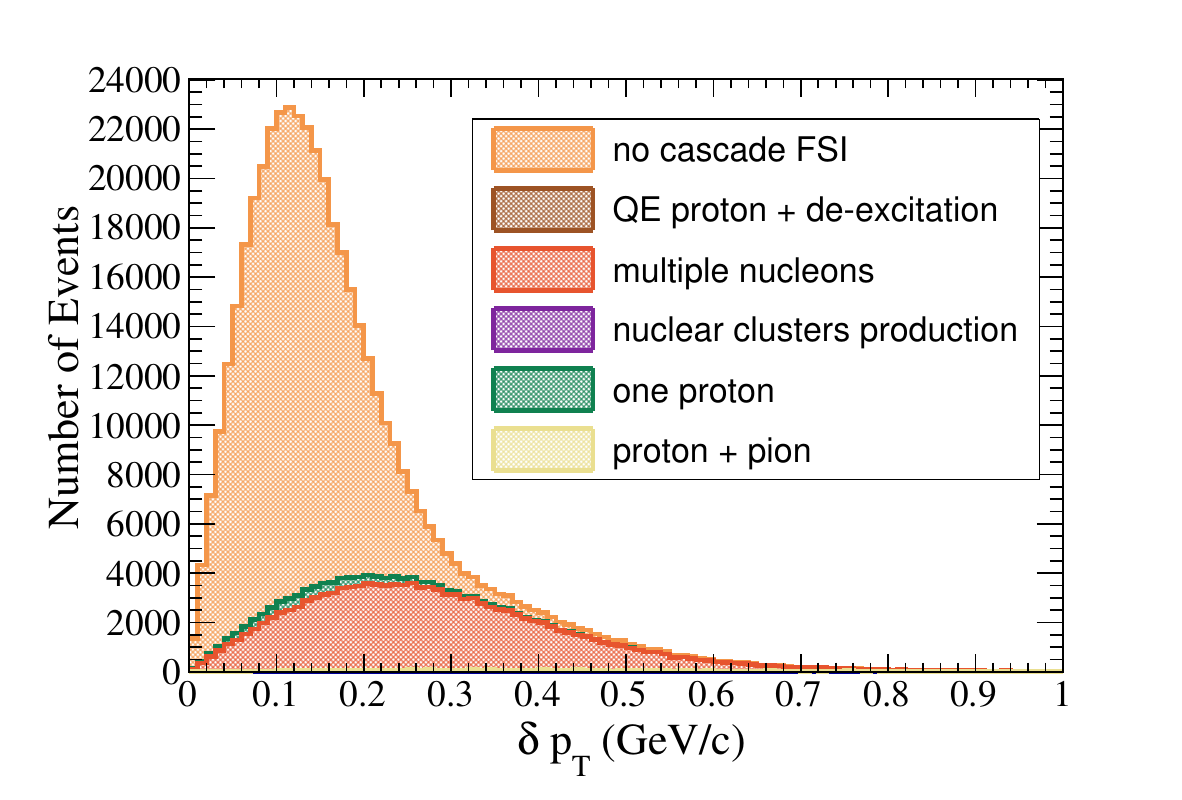} \\}
\end{minipage}
\caption{\label{fig:stv} $\dalphat$ (left) and $\dpt$ (right) simulated with INCL (top), INCL+ABLA (middle), and NuWro (bottom) models for CCQE events and T2K neutrino flux. }
\end{figure*}

Fig.~\ref{fig:stv} shows these two variables simulated with INCL+ABLA, INCL alone, and NuWro. The $\dpt$ shape is very similar between the two models since it is sensitive to the NuWro-simulated initial neutron momentum. $\dalphat$ features instead a massive difference in these two simulations in the high $\dalphat$ region. ABLA produces protons with a momentum that is mostly lower than the momentum of protons produced during a cascade. De-excitation will not change its kinematics if an event already contains the leading proton from the cascade. If, after the cascade, there was no proton in the cascade, the leading proton produced in de-excitation will contribute to high values of $\dalphat$. The very high $\dalphat$ values in INCL+ABLA simulation are constrained by the Pauli blocking suppression of the too-low momentum particles during cascade FSI, inducing a peak shape near $180^o$. 

\begin{figure}[ht]
\centering
\center{\includegraphics[width=1\linewidth]{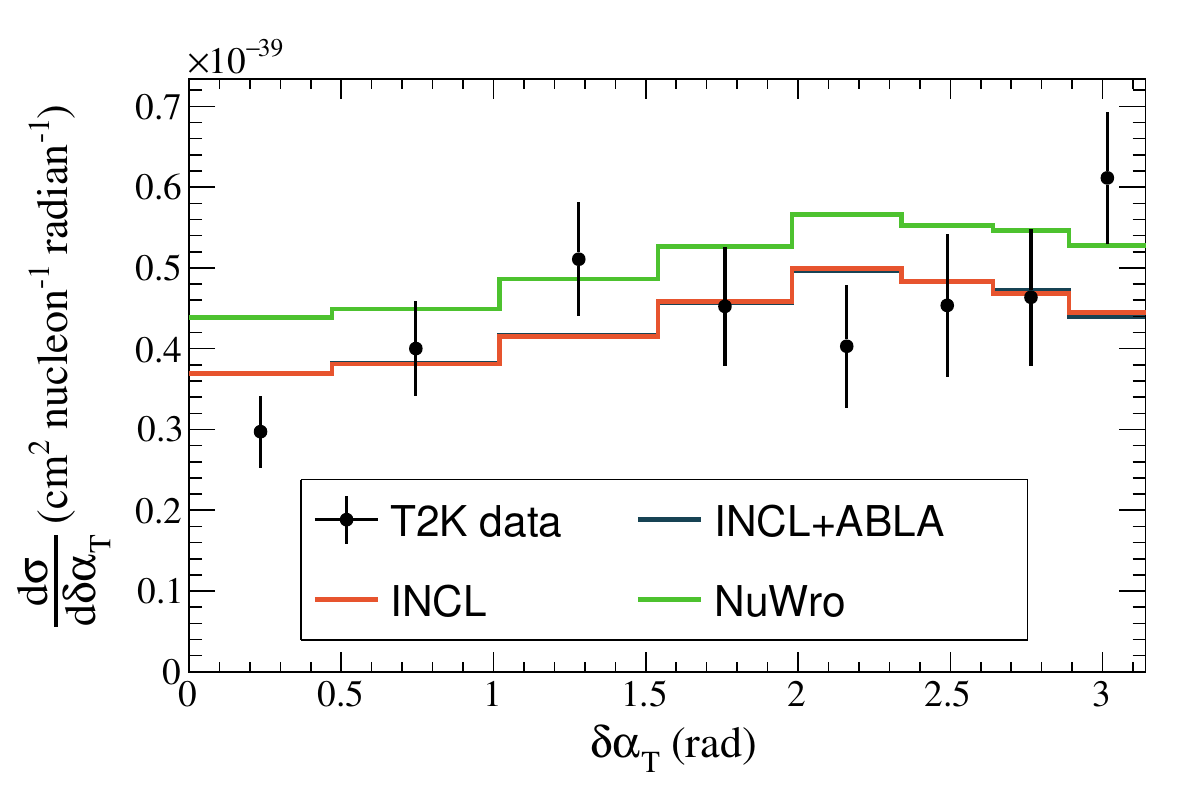} \\}
\center{\includegraphics[width=1\linewidth]{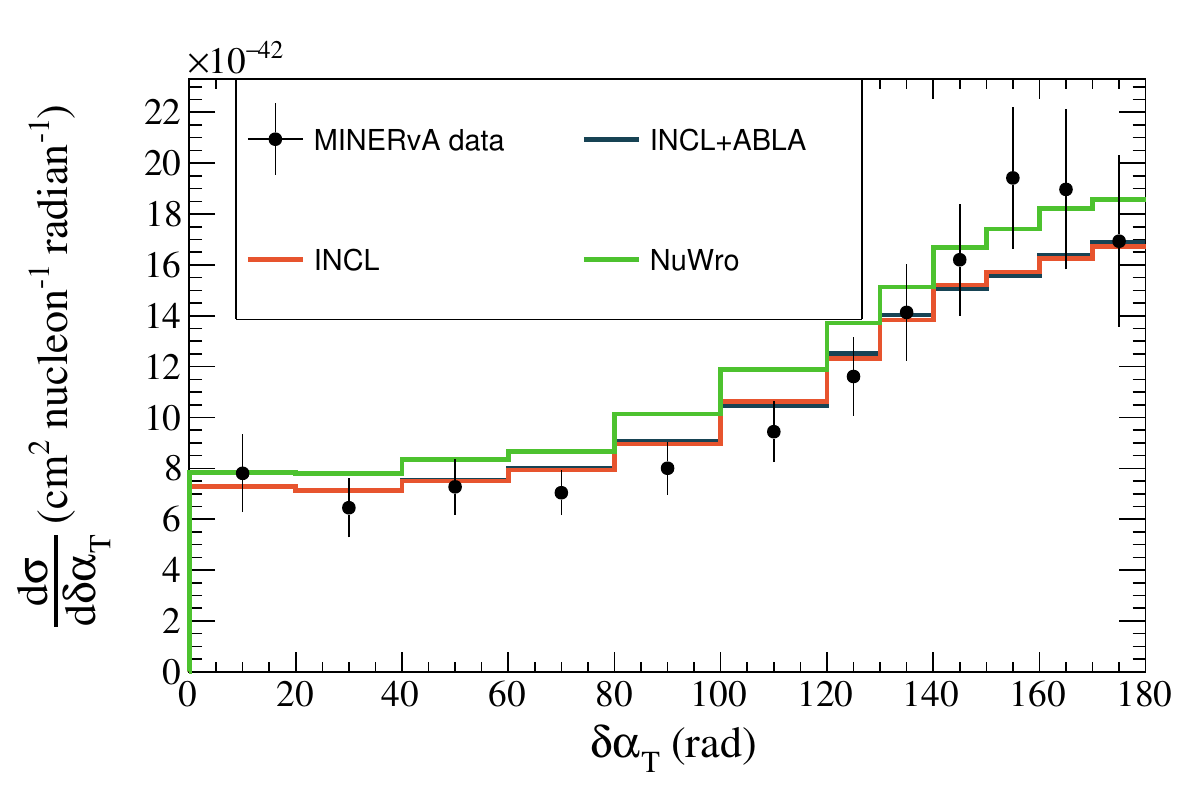} \\}
\caption{\label{fig:dataSF_T2K} $\dalphat$ simulated with INCL, INCL+ABLA and NuWro comparison to the T2K (top) and MINER$\nu$A (bottom) data.}
\end{figure}

We have compared the STV prediction of NuWro, INCL, and INCL+ABLA simulations to the T2K~\cite{T2K:2018rnz} and MINER$\nu$A~\cite{MINERvA:2018hba} data in Fig.~\ref{fig:dataSF_T2K}. We have applied kinematic cuts to simulate the detector acceptance. ABLA produces more events with at least one proton in the final state than INCL alone. On the other hand, as discussed above, ABLA produces low-momentum protons, and most of these events are rejected by acceptance cuts (in particular, proton momentum of more than 450~MeV). 

\begin{figure}[ht]
\centering
\center{\includegraphics[width=1\linewidth]{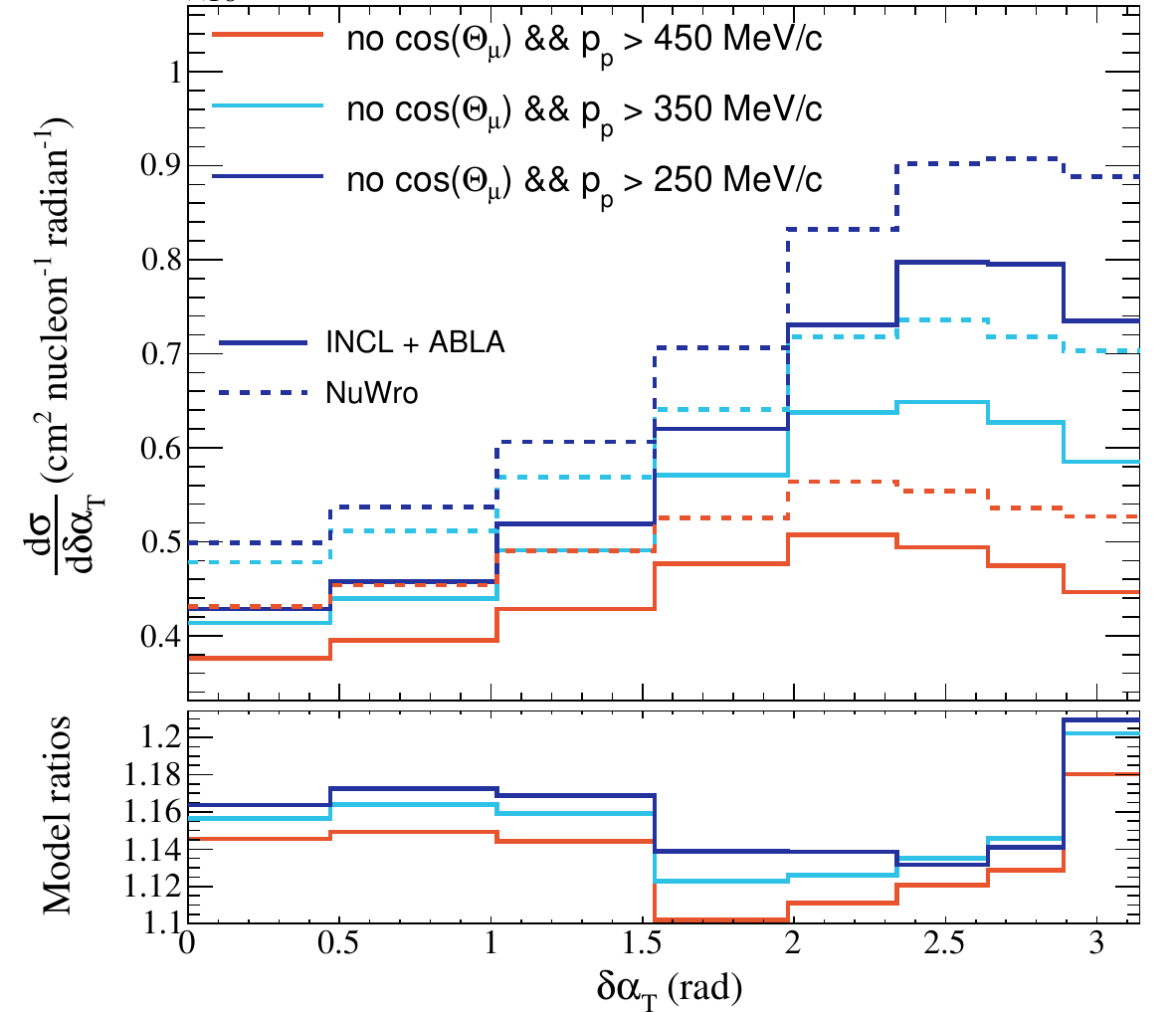} \\}
\caption{\label{fig:var_cuts} $\dalphat$ simulated with INCL+ABLA (solid lines) and NuWro (dashed lines) with no cuts on muon angle ($\Theta_\mu$) and different options of detector acceptance for the leading proton momentum threshold ($p_p$). The bottom panel presents the ratios of NuWro and INCL+ABLA models for different cuts applied.}
\end{figure}

Despite a clear difference in the proton rate in the acceptance region between the models, present data are too sparse to suggest a clear preference between the models (For the T2K comparison, $\chi^2$ for NuWro is 25.5, for INCL~--- 18.5, INCL + ABLA~--- 18.5; the number of degrees of freedom is 8. For the MINER$\nu$A comparison, $\chi^2$ for NuWro is 23.5, for INCL~--- 19.7, INCL + ABLA~--- 20.2; the number of degrees of freedom is 12.). Also, in the acceptance region with the present momentum threshold in the T2K ND280 detector, there is no clear shape difference between the models. In Fig.~\ref{fig:var_cuts}, we have varied the detector cuts to predict how distinguishable the nuclear models will be with better detector acceptance. By decreasing the proton momentum threshold, the various nuclear models' results are more and more distinct. Depending on the future data precision, with a proton momentum threshold around 200~MeV, it is possible to see differences between the models in the $\dalphat$ distribution simulation.

\begin{figure}[ht!]
\centering
\includegraphics[width=0.98\linewidth]{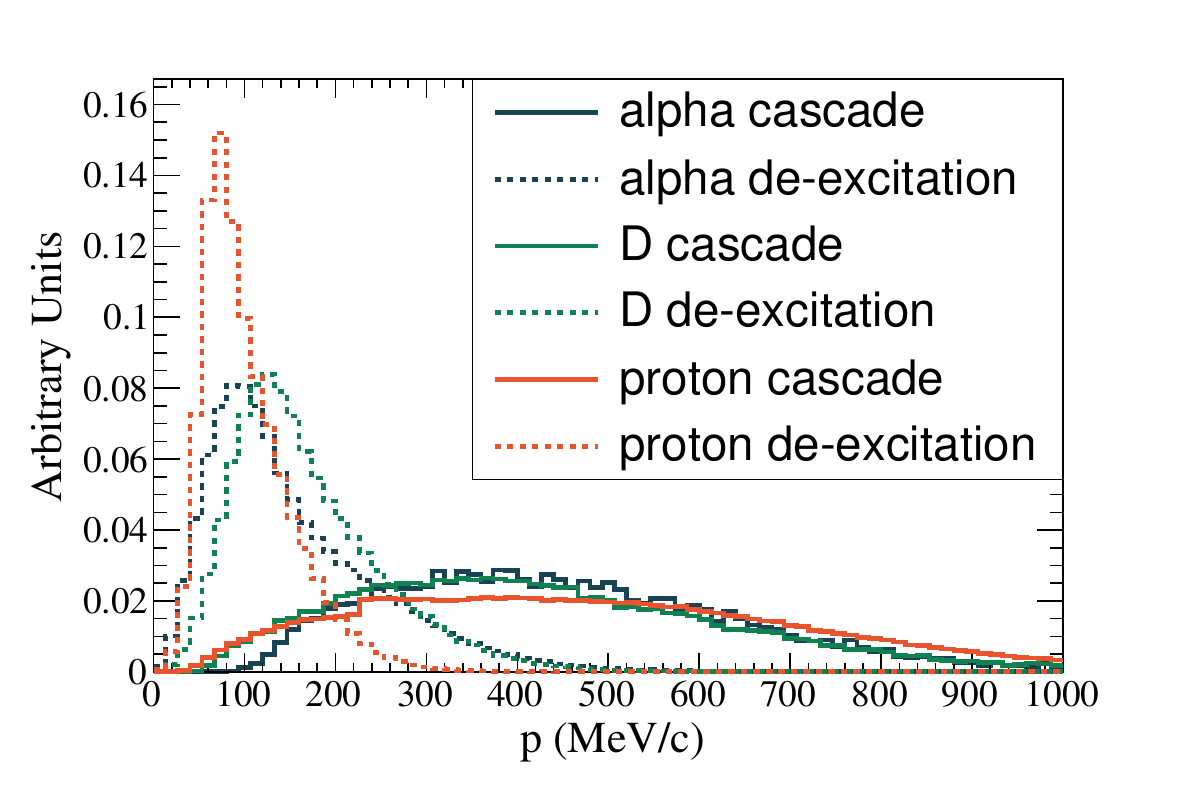}
\caption{\label{fig:mom_cascade_excit} Momentum distribution of some particles produced during cascade and de-excitation.}
\end{figure}

Fig.~\ref{fig:mom_cascade_excit} shows the momentum distribution of the most frequently produced nuclear clusters and protons emitted during the cascade and de-excitation. Multiple isotopes are produced during de-excitation, but we will focus on the most common ones: $\alpha$, deuteron, triton, and $^3$He. ABLA generates a significant amount of low-momentum particles that, in most cases, cannot leave visible tracks in the detector. 

\begin{figure}[ht!]
\centering
\includegraphics[width=0.98\linewidth]{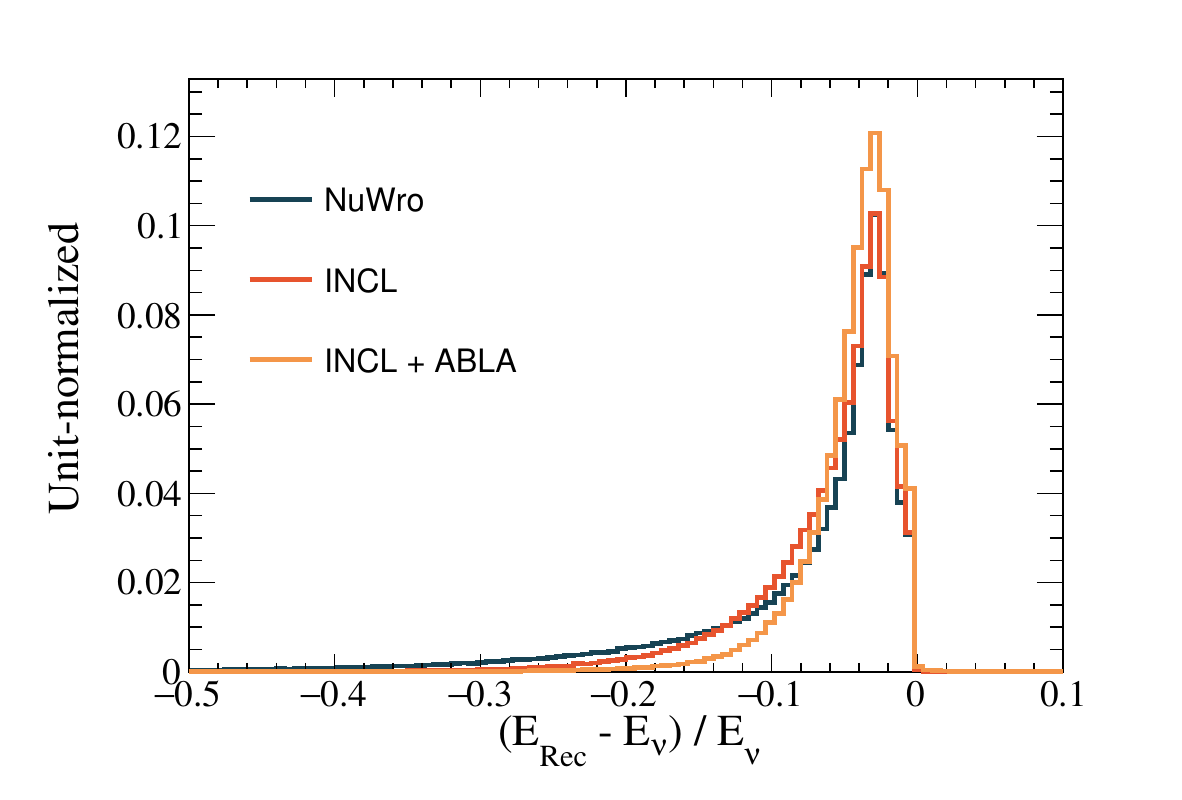}
\includegraphics[width=0.98\linewidth]{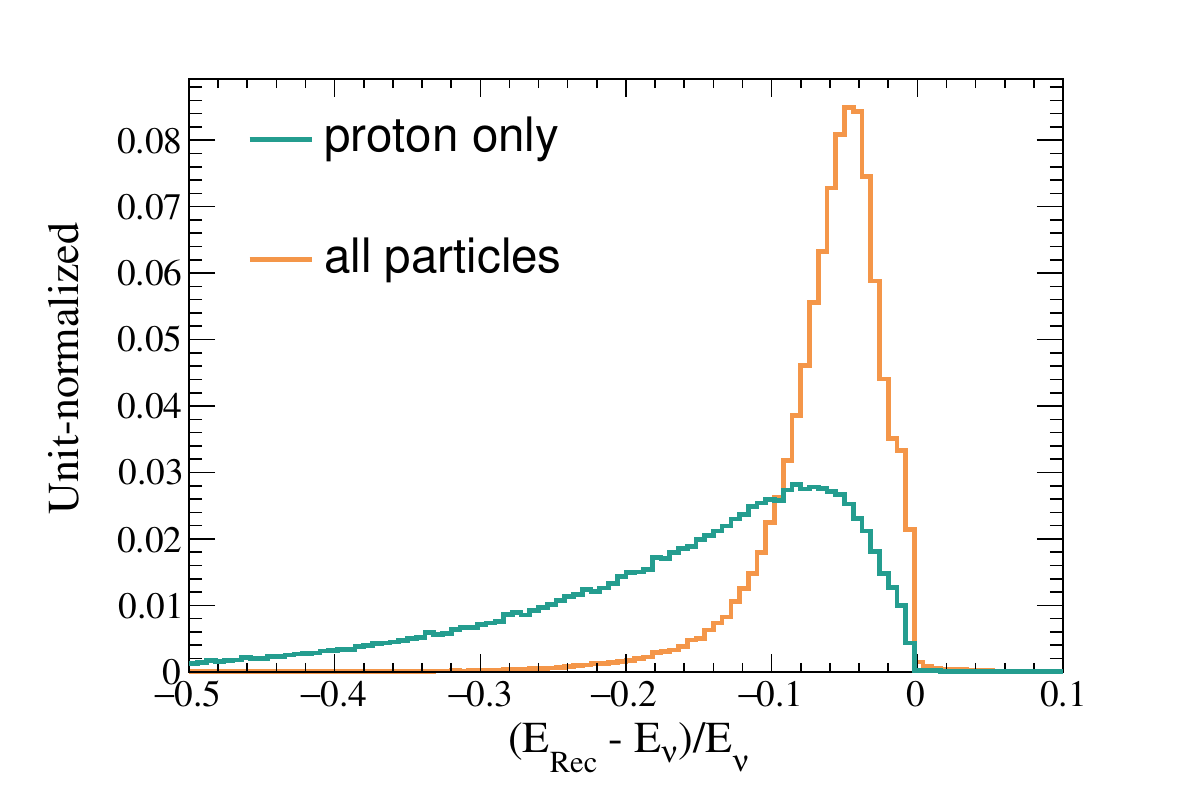}
\includegraphics[width=0.98\linewidth]{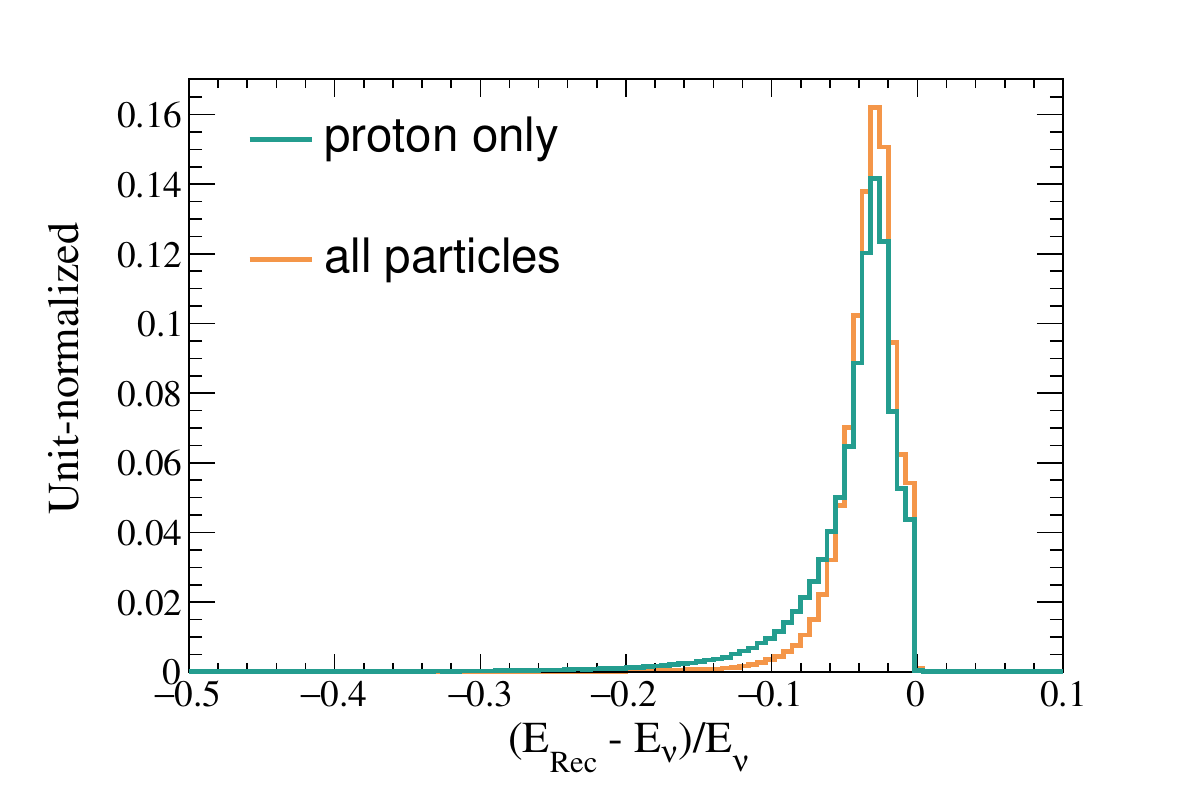}
\caption{\label{fig:abla_enurec} Shape comparison of the neutrino energy reconstruction. Top: all events simulated with NuWro, INCL, and INCL+ABLA, neutrino energy reconstruction using all outgoing particles; middle: INCL+ABLA simulation of the FSI events with proton and muon only and with all outgoing particles; bottom: INCL+ABLA simulation of the no FSI events with proton and muon only and with all outgoing particles.}
\end{figure}

We reconstruct the neutrino energy using the NuWro, INCL, and INCL+ABLA simulations. We perform reconstruction with muon and proton only and with all particles produced in the event. In Fig.~\ref{fig:abla_enurec} (top), one can see that when we reconstruct neutrino energy considering all the particles produced, the de-excitation plays an important role: the distribution obtained with INCL+ABLA is different from the ones of INCL only and NuWro. We compare the neutrino energy reconstruction using muon and proton only or all particles in Fig.~\ref{fig:abla_enurec} (middle). Including all particles, the energy reconstruction is largely improved for events where the leading proton experienced final state interactions. We enhance our energy reconstruction resolution even for no-FSI events (bottom plot of Fig.~\ref{fig:abla_enurec}) since de-excitation produces additional particles. 

\begin{figure}[ht!]
\centering
\includegraphics[width=0.98\linewidth]{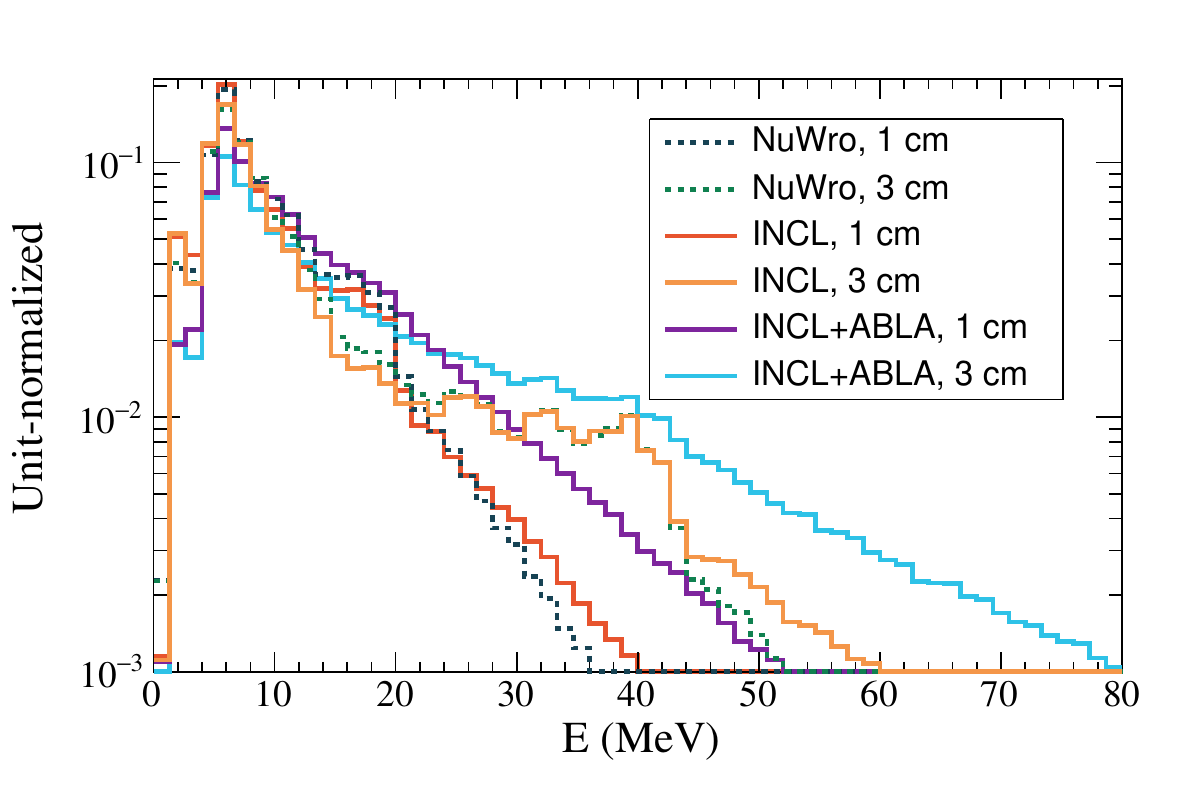}
\caption{\label{fig:VAevt} Total visible energy deposited by all particles in the event obtained with the Geant4 simulation of the CH cube in the 1(3)~cm sphere around the neutrino vertex.}
\end{figure}

Finally, to test the observability of nuclear clusters, we use Geant4~\cite{AGOSTINELLI2003250, Geant4, Allison:2016lfl} simulation to model the interaction of nuclear clusters inside a uniform hydrocarbon block. The events processed through INCL+ABLA are injected into the Geant4 simulation. Most particles will contribute to the vertex activity~--- energy deposited in a sphere around the neutrino interaction. We calculate vertex activity for the spheres with 1 and 3~cm radius around the vertex. The result of the vertex activity simulation is shown in Fig.~\ref{fig:VAevt}. We have two populations of events: particles that travel more than 1(3)~cm sphere and particles that stop inside the sphere. We apply Birks correction to simulate visible energy in the detector. The procedure is extensively described in Ref.~\cite{Ershova:2022jah}. One can observe the role of the nuclear cluster production for the vertex activity in the vertex activity predictions computed with different models. NuWro predicts 11\% of events with more than 30~MeV energy deposited around the vertex in the 3~cm sphere. For the same conditions, INCL prediction is 13\%, and INCL+ABLA is 24\%.

\begin{figure}[ht!]
\centering
\includegraphics[width=0.98\linewidth]{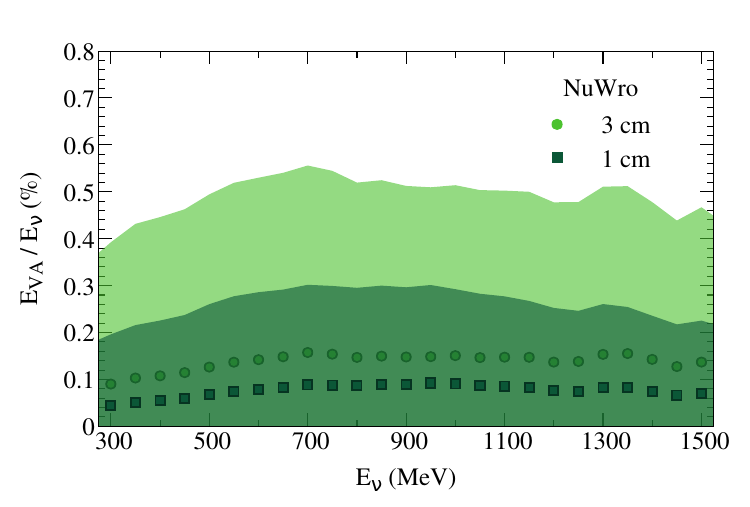}
\includegraphics[width=0.98\linewidth]{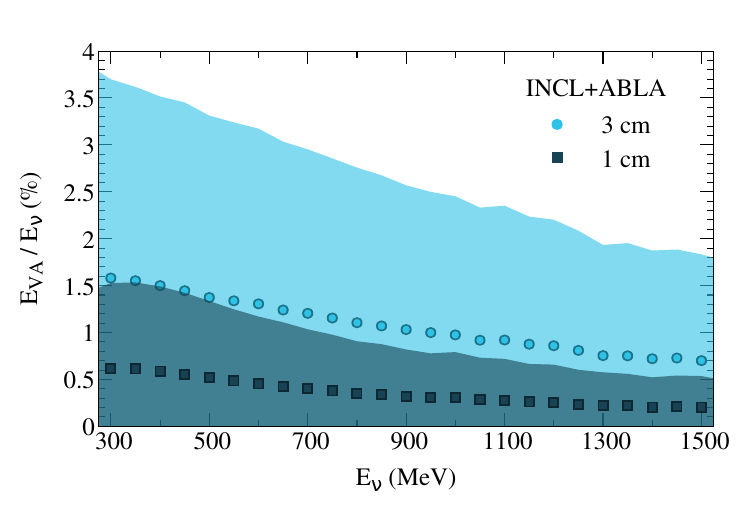}
\caption{\label{fig:VAratio} Average vertex activity as a fraction of the initial neutrino energy depending on the neutrino energy, simulated with NuWro (top) and INCL+ABLA (bottom). The leading proton and muon are not considered. The bands correspond to the standard deviation uncertainty.}
\end{figure}

\begin{figure}[ht!]
\centering
\includegraphics[width=0.98\linewidth]{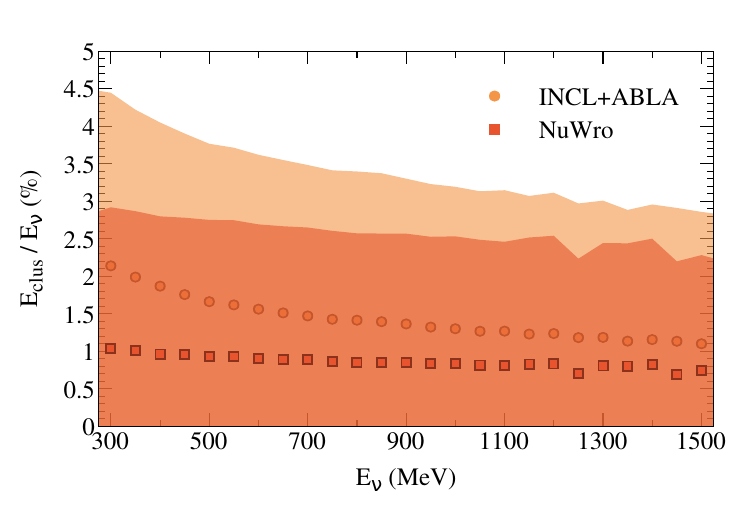}
\caption{\label{fig:ClusRatio} Average sum of the kinetic energy of all clusters, neutrons, and non-leading protons in the event as a fraction of the initial neutrino energy depending on the neutrino energy simulated with INCL+ABLA and NuWro. The bands correspond to the standard deviation uncertainty.}
\end{figure}

The larger the number of particles produced during FSI, the lower the energy of the leading proton. These particles have, in general, low momentum, causing them to deposit all their energy around the neutrino vertex primarily. As a result, the vertex activity draws energy from the leading proton. Presently, neutrino energy reconstruction does not take into account vertex activity. Fig.~\ref{fig:VAratio} illustrates the amount of neutrino energy allocated to the vertex activity (excluding the muon and leading proton) as predicted by NuWro and INCL+ABLA. The Birks correction was applied to the vertex activity calculation. Considering de-excitation, the vertex activity, on average, accounts for 1.5\% of the neutrino energy for the T2K peak neutrino energy (of about 0.6~MeV). The NuWro prediction is considerably lower: less than 0.2\% at the T2K peak neutrino energy. Fig.~\ref{fig:ClusRatio} shows the ratio of the total kinetic energy of all clusters, neutrons, and non-leading protons produced in the event ($E_{clus}$) to the true neutrino energy, representing the upper bound of the potentially detectable vertex activity. The difference shown in Fig.~\ref{fig:ClusRatio} directly corresponds to the neutrino energy reconstruction improvement for the INCL+ABLA simulation with respect to the INCL and NuWro simulations in the top panel of Fig.~\ref{fig:abla_enurec}.

\section{Conclusions}
In this paper, we have investigated the impact of final state interactions and de-excitation on neutrino-nucleus scattering modeling, focusing on the CCQE channel. By utilizing two different Monte Carlo event generators, NuWro and INCL, which is coupled with the de-excitation code ABLA, we have explored the role of FSI and nuclear de-excitation in shaping the hadronic final state, namely STV and vertex activity. To properly model the excitation energy in CCQE events, we have developed a new method of excitation energy evaluation utilizing electron scattering data analyses.

An essential novelty of this study is the release of excitation energy via the production of additional particles simulated with ABLA, which provides novel insights into the role of de-excitation in neutrino-nucleus scattering. In Ref.~\cite{Ershova:2022jah}, we have already shown and characterized nuclear clusters from the INCL cascade. Here we show that even with no re-interactions of an outgoing hadron in the nuclear medium, there is excitation energy from the primary neutrino interaction, which is later released via the production of additional particles. On average, particles produced during the de-excitation stage feature lower momentum but higher multiplicity than those produced during the cascade. In the INCL+ABLA simulation, the proton and $\alpha$ productions are enhanced by more than a factor two compared to the INCL simulation only. Such nuclear clusters, produced primarily in de-excitation, recover a fraction of the initial neutrino energy.  It is, therefore, crucial to properly model the de-excitation for precise neutrino energy reconstruction. 

Presently, modern experiments do not take into account vertex activity while reconstructing neutrino energy. In this work, we show that to reach a precision on neutrino energy reconstruction at a percent level (as requested for precise oscillation measurements), the vertex activity plays a relevant role up to several hundreds of MeV, especially when the energy released by de-excitation is considered. However, the fraction of visible energy in vertex activity (after Birks and removal of secondary interactions) tends to be lower. It is, therefore, a tough experimental challenge to measure vertex activity and correct back to the total kinetic energy of the initial particles. Hence, it is crucial to have models that can adequately describe such a fraction of energy, which needs to be corrected for a precise reconstruction of the total neutrino energy but is so difficult to observe.

In conclusion, this study has provided new insights into the role of FSI and nuclear de-excitation in neutrino-nucleus scattering. The results obtained from this study improve our understanding of nuclear effects, allowing us for designing new algorithms for neutrino energy reconstruction. This ultimately opens the road to more precise neutrino oscillation measurements.

\clearpage
\section*{Acknowledgements}
We have conducted this work within the T2K Near Detector upgrade project and gratefully acknowledge all productive meetings with our colleagues in this context. We appreciate the fruitful discussions at the INCL
meetings, particularly those with S.~Leray, D.~Mancusi, and D.~Zharenov. We are also grateful to K.~Lachner and R.~Gonz\'alez-Jim\'enez for their interest in this work and valuable comments.
This work was supported by P2IO LabEx (ANR-10-LABX-0038 – Project “BSMNu”) in the framework “Investissements d’Avenir” (ANR-11-IDEX-0003-01), managed by the Agence Nationale de la Recherche (ANR), France. We acknowledge the support of CNRS/IN2P3 and CEA. J.L.R.-S. is thankful for the support by the “Ramón y Cajal” programme under the Grant No. RYC2021-031989-I, funded by MCIN/AEI/10.13039/501100011033 and by “European Union NextGeneration-EU/PRTR.” J.T.S. and K.N. were partially supported by the Polish National Science Centre (NCN) grants No. 2021/41/B/ST2/02778 and No. 2020/37/N/ST2/01751, respectively. Additionally, K.N. acknowledges the support of the Special Research Fund, Ghent University.

\bibliography{biblio}

\providecommand{\noopsort}[1]{}\providecommand{\singleletter}[1]{#1}%
\begin{thebibliography}{66}%
\makeatletter
\providecommand \@ifxundefined [1]{%
 \@ifx{#1\undefined}
}%
\providecommand \@ifnum [1]{%
 \ifnum #1\expandafter \@firstoftwo
 \else \expandafter \@secondoftwo
 \fi
}%
\providecommand \@ifx [1]{%
 \ifx #1\expandafter \@firstoftwo
 \else \expandafter \@secondoftwo
 \fi
}%
\providecommand \natexlab [1]{#1}%
\providecommand \enquote  [1]{``#1''}%
\providecommand \bibnamefont  [1]{#1}%
\providecommand \bibfnamefont [1]{#1}%
\providecommand \citenamefont [1]{#1}%
\providecommand \href@noop [0]{\@secondoftwo}%
\providecommand \href [0]{\begingroup \@sanitize@url \@href}%
\providecommand \@href[1]{\@@startlink{#1}\@@href}%
\providecommand \@@href[1]{\endgroup#1\@@endlink}%
\providecommand \@sanitize@url [0]{\catcode `\\12\catcode `\$12\catcode
  `\&12\catcode `\#12\catcode `\^12\catcode `\_12\catcode `\%12\relax}%
\providecommand \@@startlink[1]{}%
\providecommand \@@endlink[0]{}%
\providecommand \url  [0]{\begingroup\@sanitize@url \@url }%
\providecommand \@url [1]{\endgroup\@href {#1}{\urlprefix }}%
\providecommand \urlprefix  [0]{URL }%
\providecommand \Eprint [0]{\href }%
\providecommand \doibase [0]{http://dx.doi.org/}%
\providecommand \selectlanguage [0]{\@gobble}%
\providecommand \bibinfo  [0]{\@secondoftwo}%
\providecommand \bibfield  [0]{\@secondoftwo}%
\providecommand \translation [1]{[#1]}%
\providecommand \BibitemOpen [0]{}%
\providecommand \bibitemStop [0]{}%
\providecommand \bibitemNoStop [0]{.\EOS\space}%
\providecommand \EOS [0]{\spacefactor3000\relax}%
\providecommand \BibitemShut  [1]{\csname bibitem#1\endcsname}%
\let\auto@bib@innerbib\@empty
%</preamble>
\bibitem [{\citenamefont {Ram\'\i{}rez}\ \emph {et~al.}(2023)\citenamefont
  {Ram\'\i{}rez} \emph {et~al.}}]{T2K:2023smv}%
  \BibitemOpen
  \bibfield  {author} {\bibinfo {author} {\bibfnamefont {M.~A.}\ \bibnamefont
  {Ram\'\i{}rez}} \emph {et~al.} (\bibinfo {collaboration} {T2K}),\ }\href@noop
  {} {\  (\bibinfo {year} {2023})},\ \Eprint {http://arxiv.org/abs/2303.03222}
  {arXiv:2303.03222 [hep-ex]} \BibitemShut {NoStop}%
\bibitem [{\citenamefont {Acero}\ \emph {et~al.}(2022)\citenamefont {Acero}
  \emph {et~al.}}]{Acero_2022}%
  \BibitemOpen
  \bibfield  {author} {\bibinfo {author} {\bibfnamefont {M.}~\bibnamefont
  {Acero}} \emph {et~al.},\ }\href {\doibase 10.1103/physrevd.106.032004}
  {\bibfield  {journal} {\bibinfo  {journal} {Physical Review D}\ }\textbf
  {\bibinfo {volume} {106}} (\bibinfo {year} {2022}),\
  10.1103/physrevd.106.032004}\BibitemShut {NoStop}%
\bibitem [{\citenamefont {Cabrera}\ \emph {et~al.}(2020)\citenamefont {Cabrera}
  \emph {et~al.}}]{Cabrera:2020own}%
  \BibitemOpen
  \bibfield  {author} {\bibinfo {author} {\bibfnamefont {A.}~\bibnamefont
  {Cabrera}} \emph {et~al.},\ }\href@noop {} {\  (\bibinfo {year} {2020})},\
  \Eprint {http://arxiv.org/abs/2008.11280} {arXiv:2008.11280 [hep-ph]}
  \BibitemShut {NoStop}%
\bibitem [{\citenamefont {Cao}\ \emph {et~al.}(2021)\citenamefont {Cao},
  \citenamefont {Nath}, \citenamefont {Ngoc}, \citenamefont {Francis},
  \citenamefont {Hong~Van},\ and\ \citenamefont {Quyen}}]{PhysRevD.103.112010}%
  \BibitemOpen
  \bibfield  {author} {\bibinfo {author} {\bibfnamefont {S.}~\bibnamefont
  {Cao}}, \bibinfo {author} {\bibfnamefont {A.}~\bibnamefont {Nath}}, \bibinfo
  {author} {\bibfnamefont {T.~V.}\ \bibnamefont {Ngoc}}, \bibinfo {author}
  {\bibfnamefont {N.~K.}\ \bibnamefont {Francis}}, \bibinfo {author}
  {\bibfnamefont {N.~T.}\ \bibnamefont {Hong~Van}}, \ and\ \bibinfo {author}
  {\bibfnamefont {P.~T.}\ \bibnamefont {Quyen}},\ }\href {\doibase
  10.1103/PhysRevD.103.112010} {\bibfield  {journal} {\bibinfo  {journal}
  {Phys. Rev. D}\ }\textbf {\bibinfo {volume} {103}},\ \bibinfo {pages}
  {112010} (\bibinfo {year} {2021})}\BibitemShut {NoStop}%
\bibitem [{\citenamefont {Abi}\ \emph {et~al.}(2020)\citenamefont {Abi} \emph
  {et~al.}}]{Abi_2020}%
  \BibitemOpen
  \bibfield  {author} {\bibinfo {author} {\bibfnamefont {B.}~\bibnamefont
  {Abi}} \emph {et~al.},\ }\href {\doibase 10.1088/1748-0221/15/08/T08008}
  {\bibfield  {journal} {\bibinfo  {journal} {Journal of Instrumentation}\
  }\textbf {\bibinfo {volume} {15}},\ \bibinfo {pages} {T08008} (\bibinfo
  {year} {2020})}\BibitemShut {NoStop}%
\bibitem [{\citenamefont {Abe}\ \emph {et~al.}(2018{\natexlab{a}})\citenamefont
  {Abe} \emph {et~al.}}]{Abe:2018uyc}%
  \BibitemOpen
  \bibfield  {author} {\bibinfo {author} {\bibfnamefont {K.}~\bibnamefont
  {Abe}} \emph {et~al.} (\bibinfo {collaboration} {Hyper-Kamiokande}),\
  }\href@noop {} {\  (\bibinfo {year} {2018}{\natexlab{a}})},\ \Eprint
  {http://arxiv.org/abs/1805.04163} {arXiv:1805.04163 [physics.ins-det]}
  \BibitemShut {NoStop}%
%%CITATION = ARXIV:1805.04163;%%
\bibitem [{\citenamefont {Antonello}\ \emph {et~al.}(2015)\citenamefont
  {Antonello} \emph {et~al.}}]{Antonello:2015lea}%
  \BibitemOpen
  \bibfield  {author} {\bibinfo {author} {\bibfnamefont {M.}~\bibnamefont
  {Antonello}} \emph {et~al.} (\bibinfo {collaboration} {MicroBooNE, LAr1-ND,
  ICARUS-WA104}),\ }\href@noop {} {\  (\bibinfo {year} {2015})},\ \Eprint
  {http://arxiv.org/abs/1503.01520} {arXiv:1503.01520 [physics.ins-det]}
  \BibitemShut {NoStop}%
\bibitem [{\citenamefont {Abe}\ \emph {et~al.}(2019)\citenamefont {Abe} \emph
  {et~al.}}]{T2K:2019bbb}%
  \BibitemOpen
  \bibfield  {author} {\bibinfo {author} {\bibfnamefont {K.}~\bibnamefont
  {Abe}} \emph {et~al.} (\bibinfo {collaboration} {T2K}),\ }\href@noop {} {\
  (\bibinfo {year} {2019})},\ \Eprint {http://arxiv.org/abs/1901.03750}
  {arXiv:1901.03750 [physics.ins-det]} \BibitemShut {NoStop}%
\bibitem [{\citenamefont {Alvarez-Ruso}\ \emph {et~al.}(2018)\citenamefont
  {Alvarez-Ruso} \emph {et~al.}}]{NuSTEC:2017hzk}%
  \BibitemOpen
  \bibfield  {author} {\bibinfo {author} {\bibfnamefont {L.}~\bibnamefont
  {Alvarez-Ruso}} \emph {et~al.} (\bibinfo {collaboration} {NuSTEC}),\ }\href
  {\doibase 10.1016/j.ppnp.2018.01.006} {\bibfield  {journal} {\bibinfo
  {journal} {Prog. Part. Nucl. Phys.}\ }\textbf {\bibinfo {volume} {100}},\
  \bibinfo {pages} {1} (\bibinfo {year} {2018})},\ \Eprint
  {http://arxiv.org/abs/1706.03621} {arXiv:1706.03621 [hep-ph]} \BibitemShut
  {NoStop}%
\bibitem [{\citenamefont {Martinez}\ \emph {et~al.}(2006)\citenamefont
  {Martinez}, \citenamefont {Lava}, \citenamefont {Jachowicz}, \citenamefont
  {Ryckebusch}, \citenamefont {Vantournhout},\ and\ \citenamefont
  {Udias}}]{Martinez:2005xe}%
  \BibitemOpen
  \bibfield  {author} {\bibinfo {author} {\bibfnamefont {M.~C.}\ \bibnamefont
  {Martinez}}, \bibinfo {author} {\bibfnamefont {P.}~\bibnamefont {Lava}},
  \bibinfo {author} {\bibfnamefont {N.}~\bibnamefont {Jachowicz}}, \bibinfo
  {author} {\bibfnamefont {J.}~\bibnamefont {Ryckebusch}}, \bibinfo {author}
  {\bibfnamefont {K.}~\bibnamefont {Vantournhout}}, \ and\ \bibinfo {author}
  {\bibfnamefont {J.~M.}\ \bibnamefont {Udias}},\ }\href {\doibase
  10.1103/PhysRevC.73.024607} {\bibfield  {journal} {\bibinfo  {journal} {Phys.
  Rev. C}\ }\textbf {\bibinfo {volume} {73}},\ \bibinfo {pages} {024607}
  (\bibinfo {year} {2006})},\ \Eprint {http://arxiv.org/abs/nucl-th/0505008}
  {arXiv:nucl-th/0505008} \BibitemShut {NoStop}%
\bibitem [{\citenamefont {Gonz\'alez-Jim\'enez}\ \emph
  {et~al.}(2020)\citenamefont {Gonz\'alez-Jim\'enez}, \citenamefont {Barbaro},
  \citenamefont {Caballero}, \citenamefont {Donnelly}, \citenamefont
  {Jachowicz}, \citenamefont {Megias}, \citenamefont {Niewczas}, \citenamefont
  {Nikolakopoulos},\ and\ \citenamefont
  {Ud\'\i{}as}}]{Gonzalez-Jimenez:2019ejf}%
  \BibitemOpen
  \bibfield  {author} {\bibinfo {author} {\bibfnamefont {R.}~\bibnamefont
  {Gonz\'alez-Jim\'enez}}, \bibinfo {author} {\bibfnamefont {M.~B.}\
  \bibnamefont {Barbaro}}, \bibinfo {author} {\bibfnamefont {J.~A.}\
  \bibnamefont {Caballero}}, \bibinfo {author} {\bibfnamefont {T.~W.}\
  \bibnamefont {Donnelly}}, \bibinfo {author} {\bibfnamefont {N.}~\bibnamefont
  {Jachowicz}}, \bibinfo {author} {\bibfnamefont {G.~D.}\ \bibnamefont
  {Megias}}, \bibinfo {author} {\bibfnamefont {K.}~\bibnamefont {Niewczas}},
  \bibinfo {author} {\bibfnamefont {A.}~\bibnamefont {Nikolakopoulos}}, \ and\
  \bibinfo {author} {\bibfnamefont {J.~M.}\ \bibnamefont {Ud\'\i{}as}},\ }\href
  {\doibase 10.1103/PhysRevC.101.015503} {\bibfield  {journal} {\bibinfo
  {journal} {Phys. Rev. C}\ }\textbf {\bibinfo {volume} {101}},\ \bibinfo
  {pages} {015503} (\bibinfo {year} {2020})},\ \Eprint
  {http://arxiv.org/abs/1909.07497} {arXiv:1909.07497 [nucl-th]} \BibitemShut
  {NoStop}%
\bibitem [{\citenamefont {Ershova}\ \emph {et~al.}(2022)\citenamefont {Ershova}
  \emph {et~al.}}]{Ershova:2022jah}%
  \BibitemOpen
  \bibfield  {author} {\bibinfo {author} {\bibfnamefont {A.}~\bibnamefont
  {Ershova}} \emph {et~al.},\ }\href {\doibase 10.1103/PhysRevD.106.032009}
  {\bibfield  {journal} {\bibinfo  {journal} {Phys. Rev. D}\ }\textbf {\bibinfo
  {volume} {106}},\ \bibinfo {pages} {032009} (\bibinfo {year} {2022})},\
  \Eprint {http://arxiv.org/abs/2202.10402} {arXiv:2202.10402 [hep-ph]}
  \BibitemShut {NoStop}%
\bibitem [{NuW()}]{NuWroREPO}%
  \BibitemOpen
  \href@noop {} {\enquote {\bibinfo {title} {Nuwro official repository},}\
  }\bibinfo {howpublished} {\url{https://github.com/NuWro/nuwro}}\BibitemShut
  {NoStop}%
\bibitem [{\citenamefont {Boudard}\ \emph {et~al.}(2013)\citenamefont
  {Boudard}, \citenamefont {Cugnon}, \citenamefont {David}, \citenamefont
  {Leray},\ and\ \citenamefont {Mancusi}}]{Boudard:2012wc}%
  \BibitemOpen
  \bibfield  {author} {\bibinfo {author} {\bibfnamefont {A.}~\bibnamefont
  {Boudard}}, \bibinfo {author} {\bibfnamefont {J.}~\bibnamefont {Cugnon}},
  \bibinfo {author} {\bibfnamefont {J.-C.}\ \bibnamefont {David}}, \bibinfo
  {author} {\bibfnamefont {S.}~\bibnamefont {Leray}}, \ and\ \bibinfo {author}
  {\bibfnamefont {D.}~\bibnamefont {Mancusi}},\ }\href {\doibase
  10.1103/PhysRevC.87.014606} {\bibfield  {journal} {\bibinfo  {journal} {Phys.
  Rev. C}\ }\textbf {\bibinfo {volume} {87}},\ \bibinfo {pages} {014606}
  (\bibinfo {year} {2013})},\ \Eprint {http://arxiv.org/abs/1210.3498}
  {arXiv:1210.3498 [nucl-th]} \BibitemShut {NoStop}%
\bibitem [{\citenamefont {Leray}\ \emph {et~al.}(2011)\citenamefont {Leray},
  \citenamefont {David}, \citenamefont {Khandaker}, \citenamefont {G.~Mank},
  \citenamefont {Otsuka}, \citenamefont {Filges}, \citenamefont {Gallmeier},
  \citenamefont {Konobeyev},\ and\ \citenamefont {Michel}}]{IAEA}%
  \BibitemOpen
  \bibfield  {author} {\bibinfo {author} {\bibfnamefont {S.}~\bibnamefont
  {Leray}}, \bibinfo {author} {\bibfnamefont {J.~C.}\ \bibnamefont {David}},
  \bibinfo {author} {\bibfnamefont {M.}~\bibnamefont {Khandaker}}, \bibinfo
  {author} {\bibfnamefont {A.~M.}\ \bibnamefont {G.~Mank}}, \bibinfo {author}
  {\bibfnamefont {N.}~\bibnamefont {Otsuka}}, \bibinfo {author} {\bibfnamefont
  {D.}~\bibnamefont {Filges}}, \bibinfo {author} {\bibfnamefont
  {F.}~\bibnamefont {Gallmeier}}, \bibinfo {author} {\bibfnamefont
  {A.}~\bibnamefont {Konobeyev}}, \ and\ \bibinfo {author} {\bibfnamefont
  {R.}~\bibnamefont {Michel}},\ }\href {\doibase 10.1140/epja/i2015-15068-1}
  {\bibfield  {journal} {\bibinfo  {journal} {J. Korean Phy. Soc.}\ }\textbf
  {\bibinfo {volume} {59}},\ \bibinfo {pages} {791} (\bibinfo {year}
  {2011})}\BibitemShut {NoStop}%
\bibitem [{\citenamefont {Juszczak}\ \emph {et~al.}(2005)\citenamefont
  {Juszczak}, \citenamefont {Nowak},\ and\ \citenamefont
  {Sobczyk}}]{Juszczak:2005wk}%
  \BibitemOpen
  \bibfield  {author} {\bibinfo {author} {\bibfnamefont {C.}~\bibnamefont
  {Juszczak}}, \bibinfo {author} {\bibfnamefont {J.~A.}\ \bibnamefont {Nowak}},
  \ and\ \bibinfo {author} {\bibfnamefont {J.~T.}\ \bibnamefont {Sobczyk}},\
  }\href {\doibase 10.1140/epjc/s2004-02086-9} {\bibfield  {journal} {\bibinfo
  {journal} {Eur. Phys. J.}\ }\textbf {\bibinfo {volume} {C39}},\ \bibinfo
  {pages} {195} (\bibinfo {year} {2005})}\BibitemShut {NoStop}%
%%CITATION = EPHJA,C39,195;%%
\bibitem [{\citenamefont {Thorpe}\ \emph {et~al.}(2021)\citenamefont {Thorpe},
  \citenamefont {Nowak}, \citenamefont {Niewczas}, \citenamefont {Sobczyk},\
  and\ \citenamefont {Juszczak}}]{Thorpe:2020tym}%
  \BibitemOpen
  \bibfield  {author} {\bibinfo {author} {\bibfnamefont {C.}~\bibnamefont
  {Thorpe}}, \bibinfo {author} {\bibfnamefont {J.}~\bibnamefont {Nowak}},
  \bibinfo {author} {\bibfnamefont {K.}~\bibnamefont {Niewczas}}, \bibinfo
  {author} {\bibfnamefont {J.~T.}\ \bibnamefont {Sobczyk}}, \ and\ \bibinfo
  {author} {\bibfnamefont {C.}~\bibnamefont {Juszczak}},\ }\href {\doibase
  10.1103/PhysRevC.104.035502} {\bibfield  {journal} {\bibinfo  {journal}
  {Phys. Rev. C}\ }\textbf {\bibinfo {volume} {104}},\ \bibinfo {pages}
  {035502} (\bibinfo {year} {2021})},\ \Eprint
  {http://arxiv.org/abs/2010.12361} {arXiv:2010.12361 [hep-ph]} \BibitemShut
  {NoStop}%
\bibitem [{\citenamefont {Juszczak}\ \emph {et~al.}(2006)\citenamefont
  {Juszczak}, \citenamefont {Nowak},\ and\ \citenamefont
  {Sobczyk}}]{Juszczak:2005zs}%
  \BibitemOpen
  \bibfield  {author} {\bibinfo {author} {\bibfnamefont {C.}~\bibnamefont
  {Juszczak}}, \bibinfo {author} {\bibfnamefont {J.~A.}\ \bibnamefont {Nowak}},
  \ and\ \bibinfo {author} {\bibfnamefont {J.~T.}\ \bibnamefont {Sobczyk}},\
  }\href {\doibase 10.1016/j.nuclphysbps.2006.08.069} {\bibfield  {journal}
  {\bibinfo  {journal} {Nucl. Phys. B Proc. Suppl.}\ }\textbf {\bibinfo
  {volume} {159}},\ \bibinfo {pages} {211} (\bibinfo {year} {2006})},\ \Eprint
  {http://arxiv.org/abs/hep-ph/0512365} {arXiv:hep-ph/0512365} \BibitemShut
  {NoStop}%
\bibitem [{\citenamefont {Bonus}\ \emph {et~al.}(2020)\citenamefont {Bonus},
  \citenamefont {Sobczyk}, \citenamefont {Siemaszko},\ and\ \citenamefont
  {Juszczak}}]{Bonus:2020yrd}%
  \BibitemOpen
  \bibfield  {author} {\bibinfo {author} {\bibfnamefont {T.}~\bibnamefont
  {Bonus}}, \bibinfo {author} {\bibfnamefont {J.~T.}\ \bibnamefont {Sobczyk}},
  \bibinfo {author} {\bibfnamefont {M.}~\bibnamefont {Siemaszko}}, \ and\
  \bibinfo {author} {\bibfnamefont {C.}~\bibnamefont {Juszczak}},\ }\href
  {\doibase 10.1103/PhysRevC.102.015502} {\bibfield  {journal} {\bibinfo
  {journal} {Phys. Rev. C}\ }\textbf {\bibinfo {volume} {102}},\ \bibinfo
  {pages} {015502} (\bibinfo {year} {2020})},\ \Eprint
  {http://arxiv.org/abs/2003.00088} {arXiv:2003.00088 [hep-ex]} \BibitemShut
  {NoStop}%
\bibitem [{\citenamefont {Berger}\ and\ \citenamefont
  {Sehgal}(2009)}]{Berger:2008xs}%
  \BibitemOpen
  \bibfield  {author} {\bibinfo {author} {\bibfnamefont {C.}~\bibnamefont
  {Berger}}\ and\ \bibinfo {author} {\bibfnamefont {L.~M.}\ \bibnamefont
  {Sehgal}},\ }\href {\doibase 10.1103/PhysRevD.79.053003} {\bibfield
  {journal} {\bibinfo  {journal} {Phys. Rev. D}\ }\textbf {\bibinfo {volume}
  {79}},\ \bibinfo {pages} {053003} (\bibinfo {year} {2009})},\ \Eprint
  {http://arxiv.org/abs/0812.2653} {arXiv:0812.2653 [hep-ph]} \BibitemShut
  {NoStop}%
\bibitem [{\citenamefont {Zhuridov}\ \emph {et~al.}(2021)\citenamefont
  {Zhuridov}, \citenamefont {Sobczyk}, \citenamefont {Juszczak},\ and\
  \citenamefont {Niewczas}}]{Zhuridov:2020hqu}%
  \BibitemOpen
  \bibfield  {author} {\bibinfo {author} {\bibfnamefont {D.}~\bibnamefont
  {Zhuridov}}, \bibinfo {author} {\bibfnamefont {J.~T.}\ \bibnamefont
  {Sobczyk}}, \bibinfo {author} {\bibfnamefont {C.}~\bibnamefont {Juszczak}}, \
  and\ \bibinfo {author} {\bibfnamefont {K.}~\bibnamefont {Niewczas}},\ }\href
  {\doibase 10.1088/1361-6471/abdade} {\bibfield  {journal} {\bibinfo
  {journal} {J. Phys. G}\ }\textbf {\bibinfo {volume} {48}},\ \bibinfo {pages}
  {055002} (\bibinfo {year} {2021})},\ \Eprint
  {http://arxiv.org/abs/2007.14426} {arXiv:2007.14426 [hep-ph]} \BibitemShut
  {NoStop}%
\bibitem [{\citenamefont {Golan}\ \emph {et~al.}(2012)\citenamefont {Golan},
  \citenamefont {Juszczak},\ and\ \citenamefont {Sobczyk}}]{Golan:2012wx}%
  \BibitemOpen
  \bibfield  {author} {\bibinfo {author} {\bibfnamefont {T.}~\bibnamefont
  {Golan}}, \bibinfo {author} {\bibfnamefont {C.}~\bibnamefont {Juszczak}}, \
  and\ \bibinfo {author} {\bibfnamefont {J.~T.}\ \bibnamefont {Sobczyk}},\
  }\href {\doibase 10.1103/PhysRevC.86.015505} {\bibfield  {journal} {\bibinfo
  {journal} {Phys. Rev. C}\ }\textbf {\bibinfo {volume} {86}},\ \bibinfo
  {pages} {015505} (\bibinfo {year} {2012})},\ \Eprint
  {http://arxiv.org/abs/1202.4197} {arXiv:1202.4197 [nucl-th]} \BibitemShut
  {NoStop}%
\bibitem [{\citenamefont {Niewczas}\ and\ \citenamefont
  {Sobczyk}(2019)}]{Niewczas:2019fro}%
  \BibitemOpen
  \bibfield  {author} {\bibinfo {author} {\bibfnamefont {K.}~\bibnamefont
  {Niewczas}}\ and\ \bibinfo {author} {\bibfnamefont {J.~T.}\ \bibnamefont
  {Sobczyk}},\ }\href {\doibase 10.1103/PhysRevC.100.015505} {\bibfield
  {journal} {\bibinfo  {journal} {Phys. Rev. C}\ }\textbf {\bibinfo {volume}
  {100}},\ \bibinfo {pages} {015505} (\bibinfo {year} {2019})},\ \Eprint
  {http://arxiv.org/abs/1902.05618} {arXiv:1902.05618 [hep-ex]} \BibitemShut
  {NoStop}%
\bibitem [{\citenamefont {Niewczas}\ \emph {et~al.}(2021)\citenamefont
  {Niewczas}, \citenamefont {Nikolakopoulos}, \citenamefont {Sobczyk},
  \citenamefont {Jachowicz},\ and\ \citenamefont
  {Gonz\'alez-Jim\'enez}}]{Niewczas:2020fev}%
  \BibitemOpen
  \bibfield  {author} {\bibinfo {author} {\bibfnamefont {K.}~\bibnamefont
  {Niewczas}}, \bibinfo {author} {\bibfnamefont {A.}~\bibnamefont
  {Nikolakopoulos}}, \bibinfo {author} {\bibfnamefont {J.~T.}\ \bibnamefont
  {Sobczyk}}, \bibinfo {author} {\bibfnamefont {N.}~\bibnamefont {Jachowicz}},
  \ and\ \bibinfo {author} {\bibfnamefont {R.}~\bibnamefont
  {Gonz\'alez-Jim\'enez}},\ }\href {\doibase 10.1103/PhysRevD.103.053003}
  {\bibfield  {journal} {\bibinfo  {journal} {Phys. Rev. D}\ }\textbf {\bibinfo
  {volume} {103}},\ \bibinfo {pages} {053003} (\bibinfo {year} {2021})},\
  \Eprint {http://arxiv.org/abs/2011.05269} {arXiv:2011.05269 [hep-ph]}
  \BibitemShut {NoStop}%
\bibitem [{\citenamefont {Frullani}\ and\ \citenamefont
  {Mougey}(1984)}]{Frullani:1984nn}%
  \BibitemOpen
  \bibfield  {author} {\bibinfo {author} {\bibfnamefont {S.}~\bibnamefont
  {Frullani}}\ and\ \bibinfo {author} {\bibfnamefont {J.}~\bibnamefont
  {Mougey}},\ }\href@noop {} {\bibfield  {journal} {\bibinfo  {journal} {Adv.
  Nucl. Phys.}\ }\textbf {\bibinfo {volume} {14}},\ \bibinfo {pages} {1}
  (\bibinfo {year} {1984})}\BibitemShut {NoStop}%
\bibitem [{\citenamefont {Benhar}\ \emph {et~al.}(1994)\citenamefont {Benhar},
  \citenamefont {Fabrocini}, \citenamefont {Fantoni},\ and\ \citenamefont
  {Sick}}]{BENHAR1994493}%
  \BibitemOpen
  \bibfield  {author} {\bibinfo {author} {\bibfnamefont {O.}~\bibnamefont
  {Benhar}}, \bibinfo {author} {\bibfnamefont {A.}~\bibnamefont {Fabrocini}},
  \bibinfo {author} {\bibfnamefont {S.}~\bibnamefont {Fantoni}}, \ and\
  \bibinfo {author} {\bibfnamefont {I.}~\bibnamefont {Sick}},\ }\href {\doibase
  https://doi.org/10.1016/0375-9474(94)90920-2} {\bibfield  {journal} {\bibinfo
   {journal} {Nuclear Physics A}\ }\textbf {\bibinfo {volume} {579}},\ \bibinfo
  {pages} {493} (\bibinfo {year} {1994})}\BibitemShut {NoStop}%
\bibitem [{\citenamefont {Ankowski}\ and\ \citenamefont
  {Sobczyk}(2008)}]{Ankowski:2007uy}%
  \BibitemOpen
  \bibfield  {author} {\bibinfo {author} {\bibfnamefont {A.~M.}\ \bibnamefont
  {Ankowski}}\ and\ \bibinfo {author} {\bibfnamefont {J.~T.}\ \bibnamefont
  {Sobczyk}},\ }\href {\doibase 10.1103/PhysRevC.77.044311} {\bibfield
  {journal} {\bibinfo  {journal} {Phys. Rev. C}\ }\textbf {\bibinfo {volume}
  {77}},\ \bibinfo {pages} {044311} (\bibinfo {year} {2008})},\ \Eprint
  {http://arxiv.org/abs/0711.2031} {arXiv:0711.2031 [nucl-th]} \BibitemShut
  {NoStop}%
\bibitem [{\citenamefont {Rohe}\ \emph {et~al.}(2005)\citenamefont {Rohe} \emph
  {et~al.}}]{E97-006:2005jlg}%
  \BibitemOpen
  \bibfield  {author} {\bibinfo {author} {\bibfnamefont {D.}~\bibnamefont
  {Rohe}} \emph {et~al.} (\bibinfo {collaboration} {E97-006}),\ }\href
  {\doibase 10.1103/PhysRevC.72.054602} {\bibfield  {journal} {\bibinfo
  {journal} {Phys. Rev. C}\ }\textbf {\bibinfo {volume} {72}},\ \bibinfo
  {pages} {054602} (\bibinfo {year} {2005})},\ \Eprint
  {http://arxiv.org/abs/nucl-ex/0506007} {arXiv:nucl-ex/0506007} \BibitemShut
  {NoStop}%
\bibitem [{\citenamefont {Nikolakopoulos}\ \emph {et~al.}(2019)\citenamefont
  {Nikolakopoulos}, \citenamefont {Jachowicz}, \citenamefont {Van~Dessel},
  \citenamefont {Niewczas}, \citenamefont {Gonz\'alez-Jim\'enez}, \citenamefont
  {Ud\'\i{}as},\ and\ \citenamefont {Pandey}}]{Nikolakopoulos:2019qcr}%
  \BibitemOpen
  \bibfield  {author} {\bibinfo {author} {\bibfnamefont {A.}~\bibnamefont
  {Nikolakopoulos}}, \bibinfo {author} {\bibfnamefont {N.}~\bibnamefont
  {Jachowicz}}, \bibinfo {author} {\bibfnamefont {N.}~\bibnamefont
  {Van~Dessel}}, \bibinfo {author} {\bibfnamefont {K.}~\bibnamefont
  {Niewczas}}, \bibinfo {author} {\bibfnamefont {R.}~\bibnamefont
  {Gonz\'alez-Jim\'enez}}, \bibinfo {author} {\bibfnamefont {J.~M.}\
  \bibnamefont {Ud\'\i{}as}}, \ and\ \bibinfo {author} {\bibfnamefont
  {V.}~\bibnamefont {Pandey}},\ }\href {\doibase
  10.1103/PhysRevLett.123.052501} {\bibfield  {journal} {\bibinfo  {journal}
  {Phys. Rev. Lett.}\ }\textbf {\bibinfo {volume} {123}},\ \bibinfo {pages}
  {052501} (\bibinfo {year} {2019})},\ \Eprint
  {http://arxiv.org/abs/1901.08050} {arXiv:1901.08050 [nucl-th]} \BibitemShut
  {NoStop}%
\bibitem [{\citenamefont {Ankowski}\ \emph {et~al.}(2015)\citenamefont
  {Ankowski}, \citenamefont {Benhar},\ and\ \citenamefont
  {Sakuda}}]{Ankowski:2014yfa}%
  \BibitemOpen
  \bibfield  {author} {\bibinfo {author} {\bibfnamefont {A.~M.}\ \bibnamefont
  {Ankowski}}, \bibinfo {author} {\bibfnamefont {O.}~\bibnamefont {Benhar}}, \
  and\ \bibinfo {author} {\bibfnamefont {M.}~\bibnamefont {Sakuda}},\ }\href
  {\doibase 10.1103/PhysRevD.91.033005} {\bibfield  {journal} {\bibinfo
  {journal} {Phys. Rev. D}\ }\textbf {\bibinfo {volume} {91}},\ \bibinfo
  {pages} {033005} (\bibinfo {year} {2015})},\ \Eprint
  {http://arxiv.org/abs/1404.5687} {arXiv:1404.5687 [nucl-th]} \BibitemShut
  {NoStop}%
\bibitem [{\citenamefont {Nikolakopoulos}\ \emph {et~al.}(2022)\citenamefont
  {Nikolakopoulos}, \citenamefont {Gonz\'alez-Jim\'enez}, \citenamefont
  {Jachowicz}, \citenamefont {Niewczas}, \citenamefont {S\'anchez},\ and\
  \citenamefont {Ud\'\i{}as}}]{Nikolakopoulos:2022qkq}%
  \BibitemOpen
  \bibfield  {author} {\bibinfo {author} {\bibfnamefont {A.}~\bibnamefont
  {Nikolakopoulos}}, \bibinfo {author} {\bibfnamefont {R.}~\bibnamefont
  {Gonz\'alez-Jim\'enez}}, \bibinfo {author} {\bibfnamefont {N.}~\bibnamefont
  {Jachowicz}}, \bibinfo {author} {\bibfnamefont {K.}~\bibnamefont {Niewczas}},
  \bibinfo {author} {\bibfnamefont {F.}~\bibnamefont {S\'anchez}}, \ and\
  \bibinfo {author} {\bibfnamefont {J.~M.}\ \bibnamefont {Ud\'\i{}as}},\ }\href
  {\doibase 10.1103/PhysRevC.105.054603} {\bibfield  {journal} {\bibinfo
  {journal} {Phys. Rev. C}\ }\textbf {\bibinfo {volume} {105}},\ \bibinfo
  {pages} {054603} (\bibinfo {year} {2022})},\ \Eprint
  {http://arxiv.org/abs/2202.01689} {arXiv:2202.01689 [nucl-th]} \BibitemShut
  {NoStop}%
\bibitem [{\citenamefont {David}(2015)}]{2015}%
  \BibitemOpen
  \bibfield  {author} {\bibinfo {author} {\bibfnamefont {J.~C.}\ \bibnamefont
  {David}},\ }\href {\doibase 10.1140/epja/i2015-15068-1} {\bibfield  {journal}
  {\bibinfo  {journal} {The European Physical Journal A}\ }\textbf {\bibinfo
  {volume} {51}} (\bibinfo {year} {2015}),\
  10.1140/epja/i2015-15068-1}\BibitemShut {NoStop}%
\bibitem [{\citenamefont {Rodr\'{i}guez-S\'{a}nchez}\ \emph
  {et~al.}(2022)\citenamefont {Rodr\'{i}guez-S\'{a}nchez}, \citenamefont
  {Cugnon}, \citenamefont {David}, \citenamefont {Hirtz}, \citenamefont
  {Keli\'{c}-Heil},\ and\ \citenamefont {Leray}}]{Rodriguez-Sanchez:2022Abla}%
  \BibitemOpen
  \bibfield  {author} {\bibinfo {author} {\bibfnamefont {J.~L.}\ \bibnamefont
  {Rodr\'{i}guez-S\'{a}nchez}}, \bibinfo {author} {\bibfnamefont
  {J.}~\bibnamefont {Cugnon}}, \bibinfo {author} {\bibfnamefont {J.-C.}\
  \bibnamefont {David}}, \bibinfo {author} {\bibfnamefont {J.}~\bibnamefont
  {Hirtz}}, \bibinfo {author} {\bibfnamefont {A.}~\bibnamefont
  {Keli\'{c}-Heil}}, \ and\ \bibinfo {author} {\bibfnamefont {S.}~\bibnamefont
  {Leray}},\ }\href {\doibase 10.1103/PhysRevC.105.014623} {\bibfield
  {journal} {\bibinfo  {journal} {Phys. Rev. C}\ }\textbf {\bibinfo {volume}
  {105}},\ \bibinfo {pages} {014623} (\bibinfo {year} {2022})}\BibitemShut
  {NoStop}%
\bibitem [{ABL()}]{ABLA}%
  \BibitemOpen
  \href@noop {} {\emph {\bibinfo {title} {{A. Kelić, M. V. Ricciardi, and
  K.-H. Schmidt, in Proceedings of Joint ICTP-IAEA Advanced Workshop on Model
  Codes for Spallation Reactions, ICTP Trieste, Italy, 4–8 February 2008,
  edited by D. Filges, S. Leray, Y. Yariv, A. Mengoni, A. Stanculescu, and G.
  Mank (IAEA INDC, Vienna, 2008)}}}}\BibitemShut {NoStop}%
\bibitem [{\citenamefont {Botvina}\ \emph {et~al.}(1995)\citenamefont {Botvina}
  \emph {et~al.}}]{Botvina:1994vj}%
  \BibitemOpen
  \bibfield  {author} {\bibinfo {author} {\bibfnamefont {A.~S.}\ \bibnamefont
  {Botvina}} \emph {et~al.},\ }\href {\doibase 10.1016/0375-9474(94)00621-S}
  {\bibfield  {journal} {\bibinfo  {journal} {Nucl. Phys. A}\ }\textbf
  {\bibinfo {volume} {584}},\ \bibinfo {pages} {737} (\bibinfo {year}
  {1995})}\BibitemShut {NoStop}%
\bibitem [{\citenamefont {Bondorf}\ \emph {et~al.}(1995)\citenamefont
  {Bondorf}, \citenamefont {Botvina}, \citenamefont {Ilinov}, \citenamefont
  {Mishustin},\ and\ \citenamefont {Sneppen}}]{Bondorf:1995ua}%
  \BibitemOpen
  \bibfield  {author} {\bibinfo {author} {\bibfnamefont {J.~P.}\ \bibnamefont
  {Bondorf}}, \bibinfo {author} {\bibfnamefont {A.~S.}\ \bibnamefont
  {Botvina}}, \bibinfo {author} {\bibfnamefont {A.~S.}\ \bibnamefont {Ilinov}},
  \bibinfo {author} {\bibfnamefont {I.~N.}\ \bibnamefont {Mishustin}}, \ and\
  \bibinfo {author} {\bibfnamefont {K.}~\bibnamefont {Sneppen}},\ }\href
  {\doibase 10.1016/0370-1573(94)00097-M} {\bibfield  {journal} {\bibinfo
  {journal} {Phys. Rept.}\ }\textbf {\bibinfo {volume} {257}},\ \bibinfo
  {pages} {133} (\bibinfo {year} {1995})}\BibitemShut {NoStop}%
\bibitem [{\citenamefont {Mancusi}\ \emph {et~al.}(2010)\citenamefont
  {Mancusi}, \citenamefont {Charity},\ and\ \citenamefont
  {Cugnon}}]{Mancusi:2010tg}%
  \BibitemOpen
  \bibfield  {author} {\bibinfo {author} {\bibfnamefont {D.}~\bibnamefont
  {Mancusi}}, \bibinfo {author} {\bibfnamefont {R.~J.}\ \bibnamefont
  {Charity}}, \ and\ \bibinfo {author} {\bibfnamefont {J.}~\bibnamefont
  {Cugnon}},\ }\href {\doibase 10.1103/PhysRevC.82.044610} {\bibfield
  {journal} {\bibinfo  {journal} {Phys. Rev. C}\ }\textbf {\bibinfo {volume}
  {82}},\ \bibinfo {pages} {044610} (\bibinfo {year} {2010})},\ \Eprint
  {http://arxiv.org/abs/1007.0963} {arXiv:1007.0963 [nucl-th]} \BibitemShut
  {NoStop}%
\bibitem [{\citenamefont {Charity}\ \emph {et~al.}(1988)\citenamefont {Charity}
  \emph {et~al.}}]{Charity:1988zz}%
  \BibitemOpen
  \bibfield  {author} {\bibinfo {author} {\bibfnamefont {R.~J.}\ \bibnamefont
  {Charity}} \emph {et~al.},\ }\href {\doibase 10.1016/0375-9474(88)90542-8}
  {\bibfield  {journal} {\bibinfo  {journal} {Nucl. Phys. A}\ }\textbf
  {\bibinfo {volume} {483}},\ \bibinfo {pages} {371} (\bibinfo {year}
  {1988})}\BibitemShut {NoStop}%
\bibitem [{\citenamefont {Braunn}\ \emph {et~al.}(2015)\citenamefont {Braunn},
  \citenamefont {Boudard}, \citenamefont {David} \emph {et~al.}}]{ABLAlow}%
  \BibitemOpen
  \bibfield  {author} {\bibinfo {author} {\bibfnamefont {B.}~\bibnamefont
  {Braunn}}, \bibinfo {author} {\bibfnamefont {A.}~\bibnamefont {Boudard}},
  \bibinfo {author} {\bibfnamefont {J.-C.}\ \bibnamefont {David}},  \emph
  {et~al.},\ }\href {\doibase https://doi.org/10.1140/epjp/i2015-15153-x}
  {\bibfield  {journal} {\bibinfo  {journal} {The European Physical Journal
  Plus}\ }\textbf {\bibinfo {volume} {130}},\ \bibinfo {pages} {153} (\bibinfo
  {year} {2015})}\BibitemShut {NoStop}%
\bibitem [{\citenamefont {Rodr\'\i{}guez-S\'anchez}\ \emph
  {et~al.}(2017)\citenamefont {Rodr\'\i{}guez-S\'anchez}, \citenamefont
  {David}, \citenamefont {Mancusi}, \citenamefont {Boudard}, \citenamefont
  {Cugnon},\ and\ \citenamefont {Leray}}]{Rodriguez-Sanchez:2017odk}%
  \BibitemOpen
  \bibfield  {author} {\bibinfo {author} {\bibfnamefont {J.~L.}\ \bibnamefont
  {Rodr\'\i{}guez-S\'anchez}}, \bibinfo {author} {\bibfnamefont {J.-C.}\
  \bibnamefont {David}}, \bibinfo {author} {\bibfnamefont {D.}~\bibnamefont
  {Mancusi}}, \bibinfo {author} {\bibfnamefont {A.}~\bibnamefont {Boudard}},
  \bibinfo {author} {\bibfnamefont {J.}~\bibnamefont {Cugnon}}, \ and\ \bibinfo
  {author} {\bibfnamefont {S.}~\bibnamefont {Leray}},\ }\href {\doibase
  10.1103/PhysRevC.96.054602} {\bibfield  {journal} {\bibinfo  {journal} {Phys.
  Rev. C}\ }\textbf {\bibinfo {volume} {96}},\ \bibinfo {pages} {054602}
  (\bibinfo {year} {2017})}\BibitemShut {NoStop}%
\bibitem [{\citenamefont {Mancusi}\ \emph {et~al.}(2014)\citenamefont
  {Mancusi}, \citenamefont {Boudard}, \citenamefont {Cugnon}, \citenamefont
  {David}, \citenamefont {Kaitaniemi},\ and\ \citenamefont
  {Leray}}]{Mancusi:2014fba}%
  \BibitemOpen
  \bibfield  {author} {\bibinfo {author} {\bibfnamefont {D.}~\bibnamefont
  {Mancusi}}, \bibinfo {author} {\bibfnamefont {A.}~\bibnamefont {Boudard}},
  \bibinfo {author} {\bibfnamefont {J.}~\bibnamefont {Cugnon}}, \bibinfo
  {author} {\bibfnamefont {J.-C.}\ \bibnamefont {David}}, \bibinfo {author}
  {\bibfnamefont {P.}~\bibnamefont {Kaitaniemi}}, \ and\ \bibinfo {author}
  {\bibfnamefont {S.}~\bibnamefont {Leray}},\ }\href {\doibase
  10.1103/PhysRevC.90.054602} {\bibfield  {journal} {\bibinfo  {journal} {Phys.
  Rev. C}\ }\textbf {\bibinfo {volume} {90}},\ \bibinfo {pages} {054602}
  (\bibinfo {year} {2014})},\ \Eprint {http://arxiv.org/abs/1407.7755}
  {arXiv:1407.7755 [nucl-th]} \BibitemShut {NoStop}%
\bibitem [{\citenamefont {Boudard}\ \emph {et~al.}(2002)\citenamefont
  {Boudard}, \citenamefont {Cugnon}, \citenamefont {Leray},\ and\ \citenamefont
  {Volant}}]{PhysRevC.66.044615}%
  \BibitemOpen
  \bibfield  {author} {\bibinfo {author} {\bibfnamefont {A.}~\bibnamefont
  {Boudard}}, \bibinfo {author} {\bibfnamefont {J.}~\bibnamefont {Cugnon}},
  \bibinfo {author} {\bibfnamefont {S.}~\bibnamefont {Leray}}, \ and\ \bibinfo
  {author} {\bibfnamefont {C.}~\bibnamefont {Volant}},\ }\href {\doibase
  10.1103/PhysRevC.66.044615} {\bibfield  {journal} {\bibinfo  {journal} {Phys.
  Rev. C}\ }\textbf {\bibinfo {volume} {66}},\ \bibinfo {pages} {044615}
  (\bibinfo {year} {2002})}\BibitemShut {NoStop}%
\bibitem [{\citenamefont {Pandharipande}\ and\ \citenamefont
  {Pieper}(1992)}]{Pandharipande:1992zz}%
  \BibitemOpen
  \bibfield  {author} {\bibinfo {author} {\bibfnamefont {V.~R.}\ \bibnamefont
  {Pandharipande}}\ and\ \bibinfo {author} {\bibfnamefont {S.~C.}\ \bibnamefont
  {Pieper}},\ }\href {\doibase 10.1103/PhysRevC.45.791} {\bibfield  {journal}
  {\bibinfo  {journal} {Phys. Rev. C}\ }\textbf {\bibinfo {volume} {45}},\
  \bibinfo {pages} {791} (\bibinfo {year} {1992})}\BibitemShut {NoStop}%
\bibitem [{\citenamefont {Weisskopf}\ and\ \citenamefont
  {Ewing}(1940)}]{Weisskopf:1940}%
  \BibitemOpen
  \bibfield  {author} {\bibinfo {author} {\bibfnamefont {V.~F.}\ \bibnamefont
  {Weisskopf}}\ and\ \bibinfo {author} {\bibfnamefont {D.~H.}\ \bibnamefont
  {Ewing}},\ }\href {\doibase 10.1103/PhysRev.57.472} {\bibfield  {journal}
  {\bibinfo  {journal} {Phys. Rev.}\ }\textbf {\bibinfo {volume} {57}},\
  \bibinfo {pages} {472} (\bibinfo {year} {1940})}\BibitemShut {NoStop}%
\bibitem [{\citenamefont {Gilbert}\ and\ \citenamefont
  {Cameron}(1965)}]{Gilbert:1965}%
  \BibitemOpen
  \bibfield  {author} {\bibinfo {author} {\bibfnamefont {A.}~\bibnamefont
  {Gilbert}}\ and\ \bibinfo {author} {\bibfnamefont {A.~G.~W.}\ \bibnamefont
  {Cameron}},\ }\href {\doibase 10.1139/p65-139} {\bibfield  {journal}
  {\bibinfo  {journal} {Canadian Journal of Physics}\ }\textbf {\bibinfo
  {volume} {43}},\ \bibinfo {pages} {1446} (\bibinfo {year} {1965})},\ \Eprint
  {http://arxiv.org/abs/https://doi.org/10.1139/p65-139}
  {https://doi.org/10.1139/p65-139} \BibitemShut {NoStop}%
\bibitem [{\citenamefont {Bethe}(1937)}]{Bethe:1937}%
  \BibitemOpen
  \bibfield  {author} {\bibinfo {author} {\bibfnamefont {H.~A.}\ \bibnamefont
  {Bethe}},\ }\href {\doibase 10.1103/RevModPhys.9.69} {\bibfield  {journal}
  {\bibinfo  {journal} {Rev. Mod. Phys.}\ }\textbf {\bibinfo {volume} {9}},\
  \bibinfo {pages} {69} (\bibinfo {year} {1937})}\BibitemShut {NoStop}%
\bibitem [{\citenamefont {Moller}\ \emph {et~al.}(1995)\citenamefont {Moller},
  \citenamefont {Nix}, \citenamefont {Myers},\ and\ \citenamefont
  {Swiatecki}}]{Moller:1995}%
  \BibitemOpen
  \bibfield  {author} {\bibinfo {author} {\bibfnamefont {P.}~\bibnamefont
  {Moller}}, \bibinfo {author} {\bibfnamefont {J.}~\bibnamefont {Nix}},
  \bibinfo {author} {\bibfnamefont {W.}~\bibnamefont {Myers}}, \ and\ \bibinfo
  {author} {\bibfnamefont {W.}~\bibnamefont {Swiatecki}},\ }\href {\doibase
  https://doi.org/10.1006/adnd.1995.1002} {\bibfield  {journal} {\bibinfo
  {journal} {Atomic Data and Nuclear Data Tables}\ }\textbf {\bibinfo {volume}
  {59}},\ \bibinfo {pages} {185} (\bibinfo {year} {1995})}\BibitemShut
  {NoStop}%
\bibitem [{\citenamefont {Junghans}\ \emph {et~al.}(1998)\citenamefont
  {Junghans}, \citenamefont {{de Jong}}, \citenamefont {Clerc}, \citenamefont
  {Ignatyuk}, \citenamefont {Kudyaev},\ and\ \citenamefont
  {Schmidt}}]{Junghans:1998}%
  \BibitemOpen
  \bibfield  {author} {\bibinfo {author} {\bibfnamefont {A.}~\bibnamefont
  {Junghans}}, \bibinfo {author} {\bibfnamefont {M.}~\bibnamefont {{de Jong}}},
  \bibinfo {author} {\bibfnamefont {H.-G.}\ \bibnamefont {Clerc}}, \bibinfo
  {author} {\bibfnamefont {A.}~\bibnamefont {Ignatyuk}}, \bibinfo {author}
  {\bibfnamefont {G.}~\bibnamefont {Kudyaev}}, \ and\ \bibinfo {author}
  {\bibfnamefont {K.-H.}\ \bibnamefont {Schmidt}},\ }\href {\doibase
  https://doi.org/10.1016/S0375-9474(98)00658-7} {\bibfield  {journal}
  {\bibinfo  {journal} {Nuclear Physics A}\ }\textbf {\bibinfo {volume}
  {629}},\ \bibinfo {pages} {635} (\bibinfo {year} {1998})}\BibitemShut
  {NoStop}%
\bibitem [{\citenamefont {Huang}\ \emph {et~al.}(2017)\citenamefont {Huang},
  \citenamefont {Audi}, \citenamefont {Meng}, \citenamefont {Kondev},
  \citenamefont {Naimi},\ and\ \citenamefont {Xing}}]{Huang:2017}%
  \BibitemOpen
  \bibfield  {author} {\bibinfo {author} {\bibfnamefont {W.}~\bibnamefont
  {Huang}}, \bibinfo {author} {\bibfnamefont {G.}~\bibnamefont {Audi}},
  \bibinfo {author} {\bibfnamefont {W.}~\bibnamefont {Meng}}, \bibinfo {author}
  {\bibfnamefont {F.~G.}\ \bibnamefont {Kondev}}, \bibinfo {author}
  {\bibfnamefont {S.}~\bibnamefont {Naimi}}, \ and\ \bibinfo {author}
  {\bibfnamefont {X.}~\bibnamefont {Xing}},\ }\href {\doibase
  10.1088/1674-1137/41/3/030002} {\bibfield  {journal} {\bibinfo  {journal}
  {Chinese Physics C}\ }\textbf {\bibinfo {volume} {41}},\ \bibinfo {pages}
  {030002} (\bibinfo {year} {2017})}\BibitemShut {NoStop}%
\bibitem [{\citenamefont {Qu}\ \emph {et~al.}(2011)\citenamefont {Qu},
  \citenamefont {Zhang},\ and\ \citenamefont {Le}}]{QU:2011}%
  \BibitemOpen
  \bibfield  {author} {\bibinfo {author} {\bibfnamefont {W.}~\bibnamefont
  {Qu}}, \bibinfo {author} {\bibfnamefont {G.}~\bibnamefont {Zhang}}, \ and\
  \bibinfo {author} {\bibfnamefont {X.}~\bibnamefont {Le}},\ }\href {\doibase
  https://doi.org/10.1016/j.nuclphysa.2011.08.002} {\bibfield  {journal}
  {\bibinfo  {journal} {Nuclear Physics A}\ }\textbf {\bibinfo {volume}
  {868-869}},\ \bibinfo {pages} {1} (\bibinfo {year} {2011})}\BibitemShut
  {NoStop}%
\bibitem [{\citenamefont {Axel}(1962)}]{Axel:1962}%
  \BibitemOpen
  \bibfield  {author} {\bibinfo {author} {\bibfnamefont {P.}~\bibnamefont
  {Axel}},\ }\href {\doibase 10.1103/PhysRev.126.671} {\bibfield  {journal}
  {\bibinfo  {journal} {Phys. Rev.}\ }\textbf {\bibinfo {volume} {126}},\
  \bibinfo {pages} {671} (\bibinfo {year} {1962})}\BibitemShut {NoStop}%
\bibitem [{Ign(2001)}]{Ignatyuk:2001}%
  \BibitemOpen
  \href@noop {} {\emph {\bibinfo {title} {Proceedings of the Conference Bologna
  2000: Structure of the Nucleus at the Dawn of the Century}}}\ (\bibinfo
  {year} {2001})\BibitemShut {NoStop}%
\bibitem [{\citenamefont {Ricciardi}\ \emph {et~al.}(2004)\citenamefont
  {Ricciardi}, \citenamefont {Ignatyuk}, \citenamefont {Kelić}, \citenamefont
  {Napolitani}, \citenamefont {Rejmund}, \citenamefont {Schmidt},\ and\
  \citenamefont {Yordanov}}]{Ricciardi:2004}%
  \BibitemOpen
  \bibfield  {author} {\bibinfo {author} {\bibfnamefont {M.}~\bibnamefont
  {Ricciardi}}, \bibinfo {author} {\bibfnamefont {A.}~\bibnamefont {Ignatyuk}},
  \bibinfo {author} {\bibfnamefont {A.}~\bibnamefont {Kelić}}, \bibinfo
  {author} {\bibfnamefont {P.}~\bibnamefont {Napolitani}}, \bibinfo {author}
  {\bibfnamefont {F.}~\bibnamefont {Rejmund}}, \bibinfo {author} {\bibfnamefont
  {K.-H.}\ \bibnamefont {Schmidt}}, \ and\ \bibinfo {author} {\bibfnamefont
  {O.}~\bibnamefont {Yordanov}},\ }\href {\doibase
  https://doi.org/10.1016/j.nuclphysa.2004.01.069} {\bibfield  {journal}
  {\bibinfo  {journal} {Nuclear Physics A}\ }\textbf {\bibinfo {volume}
  {733}},\ \bibinfo {pages} {299} (\bibinfo {year} {2004})}\BibitemShut
  {NoStop}%
\bibitem [{\citenamefont {Capote}\ \emph {et~al.}(2009)\citenamefont {Capote},
  \citenamefont {Herman}, \citenamefont {Obložinský}, \citenamefont {Young},
  \citenamefont {Goriely}, \citenamefont {Belgya}, \citenamefont {Ignatyuk},
  \citenamefont {Koning}, \citenamefont {Hilaire}, \citenamefont {Plujko},
  \citenamefont {Avrigeanu}, \citenamefont {Bersillon}, \citenamefont
  {Chadwick}, \citenamefont {Fukahori}, \citenamefont {Ge}, \citenamefont
  {Han}, \citenamefont {Kailas}, \citenamefont {Kopecky}, \citenamefont
  {Maslov}, \citenamefont {Reffo}, \citenamefont {Sin}, \citenamefont
  {Soukhovitskii},\ and\ \citenamefont {Talou}}]{Capote:2009}%
  \BibitemOpen
  \bibfield  {author} {\bibinfo {author} {\bibfnamefont {R.}~\bibnamefont
  {Capote}}, \bibinfo {author} {\bibfnamefont {M.}~\bibnamefont {Herman}},
  \bibinfo {author} {\bibfnamefont {P.}~\bibnamefont {Obložinský}}, \bibinfo
  {author} {\bibfnamefont {P.}~\bibnamefont {Young}}, \bibinfo {author}
  {\bibfnamefont {S.}~\bibnamefont {Goriely}}, \bibinfo {author} {\bibfnamefont
  {T.}~\bibnamefont {Belgya}}, \bibinfo {author} {\bibfnamefont
  {A.}~\bibnamefont {Ignatyuk}}, \bibinfo {author} {\bibfnamefont
  {A.}~\bibnamefont {Koning}}, \bibinfo {author} {\bibfnamefont
  {S.}~\bibnamefont {Hilaire}}, \bibinfo {author} {\bibfnamefont
  {V.}~\bibnamefont {Plujko}}, \bibinfo {author} {\bibfnamefont
  {M.}~\bibnamefont {Avrigeanu}}, \bibinfo {author} {\bibfnamefont
  {O.}~\bibnamefont {Bersillon}}, \bibinfo {author} {\bibfnamefont
  {M.}~\bibnamefont {Chadwick}}, \bibinfo {author} {\bibfnamefont
  {T.}~\bibnamefont {Fukahori}}, \bibinfo {author} {\bibfnamefont
  {Z.}~\bibnamefont {Ge}}, \bibinfo {author} {\bibfnamefont {Y.}~\bibnamefont
  {Han}}, \bibinfo {author} {\bibfnamefont {S.}~\bibnamefont {Kailas}},
  \bibinfo {author} {\bibfnamefont {J.}~\bibnamefont {Kopecky}}, \bibinfo
  {author} {\bibfnamefont {V.}~\bibnamefont {Maslov}}, \bibinfo {author}
  {\bibfnamefont {G.}~\bibnamefont {Reffo}}, \bibinfo {author} {\bibfnamefont
  {M.}~\bibnamefont {Sin}}, \bibinfo {author} {\bibfnamefont {E.}~\bibnamefont
  {Soukhovitskii}}, \ and\ \bibinfo {author} {\bibfnamefont {P.}~\bibnamefont
  {Talou}},\ }\href {\doibase https://doi.org/10.1016/j.nds.2009.10.004}
  {\bibfield  {journal} {\bibinfo  {journal} {Nuclear Data Sheets}\ }\textbf
  {\bibinfo {volume} {110}},\ \bibinfo {pages} {3107} (\bibinfo {year}
  {2009})},\ \bibinfo {note} {special Issue on Nuclear Reaction
  Data}\BibitemShut {NoStop}%
\bibitem [{\citenamefont {Abe}\ \emph {et~al.}(2023)\citenamefont {Abe} \emph
  {et~al.}}]{KamLAND:2022ptk}%
  \BibitemOpen
  \bibfield  {author} {\bibinfo {author} {\bibfnamefont {S.}~\bibnamefont
  {Abe}} \emph {et~al.} (\bibinfo {collaboration} {KamLAND}),\ }\href {\doibase
  10.1103/PhysRevD.107.072006} {\bibfield  {journal} {\bibinfo  {journal}
  {Phys. Rev. D}\ }\textbf {\bibinfo {volume} {107}},\ \bibinfo {pages}
  {072006} (\bibinfo {year} {2023})},\ \Eprint
  {http://arxiv.org/abs/2211.13911} {arXiv:2211.13911 [hep-ex]} \BibitemShut
  {NoStop}%
\bibitem [{\citenamefont {Mougey}\ \emph {et~al.}(1976)\citenamefont {Mougey},
  \citenamefont {Bernheim}, \citenamefont {Bussière}, \citenamefont
  {Gillebert}, \citenamefont {{Phan Xuan Hô}}, \citenamefont {Priou},
  \citenamefont {Royer}, \citenamefont {Sick},\ and\ \citenamefont
  {Wagner}}]{MOUGEY1976461}%
  \BibitemOpen
  \bibfield  {author} {\bibinfo {author} {\bibfnamefont {J.}~\bibnamefont
  {Mougey}}, \bibinfo {author} {\bibfnamefont {M.}~\bibnamefont {Bernheim}},
  \bibinfo {author} {\bibfnamefont {A.}~\bibnamefont {Bussière}}, \bibinfo
  {author} {\bibfnamefont {A.}~\bibnamefont {Gillebert}}, \bibinfo {author}
  {\bibnamefont {{Phan Xuan Hô}}}, \bibinfo {author} {\bibfnamefont
  {M.}~\bibnamefont {Priou}}, \bibinfo {author} {\bibfnamefont
  {D.}~\bibnamefont {Royer}}, \bibinfo {author} {\bibfnamefont
  {I.}~\bibnamefont {Sick}}, \ and\ \bibinfo {author} {\bibfnamefont
  {G.}~\bibnamefont {Wagner}},\ }\href {\doibase
  https://doi.org/10.1016/0375-9474(76)90510-8} {\bibfield  {journal} {\bibinfo
   {journal} {Nuclear Physics A}\ }\textbf {\bibinfo {volume} {262}},\ \bibinfo
  {pages} {461} (\bibinfo {year} {1976})}\BibitemShut {NoStop}%
\bibitem [{\citenamefont {de~Witt~Huberts}(1990)}]{deWittHuberts_1990}%
  \BibitemOpen
  \bibfield  {author} {\bibinfo {author} {\bibfnamefont {P.~K.~A.}\
  \bibnamefont {de~Witt~Huberts}},\ }\href {\doibase
  10.1088/0954-3899/16/4/004} {\bibfield  {journal} {\bibinfo  {journal}
  {Journal of Physics G: Nuclear and Particle Physics}\ }\textbf {\bibinfo
  {volume} {16}},\ \bibinfo {pages} {507} (\bibinfo {year} {1990})}\BibitemShut
  {NoStop}%
\bibitem [{\citenamefont {Van Der~Steenhoven}\ \emph
  {et~al.}(1988)\citenamefont {Van Der~Steenhoven}, \citenamefont {Blok},
  \citenamefont {Jans}, \citenamefont {Lapikas}, \citenamefont {Quint},\ and\
  \citenamefont {De~Witt~Huberts}}]{VanDerSteenhoven:1988um}%
  \BibitemOpen
  \bibfield  {author} {\bibinfo {author} {\bibfnamefont {G.}~\bibnamefont {Van
  Der~Steenhoven}}, \bibinfo {author} {\bibfnamefont {H.~P.}\ \bibnamefont
  {Blok}}, \bibinfo {author} {\bibfnamefont {E.}~\bibnamefont {Jans}}, \bibinfo
  {author} {\bibfnamefont {L.}~\bibnamefont {Lapikas}}, \bibinfo {author}
  {\bibfnamefont {E.~N.~M.}\ \bibnamefont {Quint}}, \ and\ \bibinfo {author}
  {\bibfnamefont {P.~K.~A.}\ \bibnamefont {De~Witt~Huberts}},\ }\href {\doibase
  10.1016/0375-9474(88)90304-1} {\bibfield  {journal} {\bibinfo  {journal}
  {Nucl. Phys. A}\ }\textbf {\bibinfo {volume} {484}},\ \bibinfo {pages} {445}
  (\bibinfo {year} {1988})}\BibitemShut {NoStop}%
\bibitem [{\citenamefont {Abe}(2021)}]{Abe:2021cgz}%
  \BibitemOpen
  \bibfield  {author} {\bibinfo {author} {\bibfnamefont {S.}~\bibnamefont
  {Abe}} (\bibinfo {collaboration} {KamLAND}),\ }\href {\doibase
  10.1088/1742-6596/2156/1/012189} {\bibfield  {journal} {\bibinfo  {journal}
  {J. Phys. Conf. Ser.}\ }\textbf {\bibinfo {volume} {2156}},\ \bibinfo {pages}
  {012189} (\bibinfo {year} {2021})}\BibitemShut {NoStop}%
\bibitem [{\citenamefont {Abe}\ \emph {et~al.}(2015)\citenamefont {Abe} \emph
  {et~al.}}]{T2K:2015sqm}%
  \BibitemOpen
  \bibfield  {author} {\bibinfo {author} {\bibfnamefont {K.}~\bibnamefont
  {Abe}} \emph {et~al.} (\bibinfo {collaboration} {T2K}),\ }\href {\doibase
  10.1103/PhysRevD.91.072010} {\bibfield  {journal} {\bibinfo  {journal} {Phys.
  Rev. D}\ }\textbf {\bibinfo {volume} {91}},\ \bibinfo {pages} {072010}
  (\bibinfo {year} {2015})},\ \Eprint {http://arxiv.org/abs/1502.01550}
  {arXiv:1502.01550 [hep-ex]} \BibitemShut {NoStop}%
\bibitem [{\citenamefont {Lu}\ \emph {et~al.}(2016)\citenamefont {Lu},
  \citenamefont {Pickering}, \citenamefont {Dolan}, \citenamefont {Barr},
  \citenamefont {Coplowe}, \citenamefont {Uchida}, \citenamefont {Wark},
  \citenamefont {Wascko}, \citenamefont {Weber},\ and\ \citenamefont
  {Yuan}}]{Lu:2015tcr}%
  \BibitemOpen
  \bibfield  {author} {\bibinfo {author} {\bibfnamefont {X.~G.}\ \bibnamefont
  {Lu}}, \bibinfo {author} {\bibfnamefont {L.}~\bibnamefont {Pickering}},
  \bibinfo {author} {\bibfnamefont {S.}~\bibnamefont {Dolan}}, \bibinfo
  {author} {\bibfnamefont {G.}~\bibnamefont {Barr}}, \bibinfo {author}
  {\bibfnamefont {D.}~\bibnamefont {Coplowe}}, \bibinfo {author} {\bibfnamefont
  {Y.}~\bibnamefont {Uchida}}, \bibinfo {author} {\bibfnamefont
  {D.}~\bibnamefont {Wark}}, \bibinfo {author} {\bibfnamefont {M.~O.}\
  \bibnamefont {Wascko}}, \bibinfo {author} {\bibfnamefont {A.}~\bibnamefont
  {Weber}}, \ and\ \bibinfo {author} {\bibfnamefont {T.}~\bibnamefont {Yuan}},\
  }\href {\doibase 10.1103/PhysRevC.94.015503} {\bibfield  {journal} {\bibinfo
  {journal} {Phys. Rev.}\ }\textbf {\bibinfo {volume} {C94}},\ \bibinfo {pages}
  {015503} (\bibinfo {year} {2016})},\ \Eprint
  {http://arxiv.org/abs/1512.05748} {arXiv:1512.05748 [nucl-th]} \BibitemShut
  {NoStop}%
%%CITATION = ARXIV:1512.05748;%%
\bibitem [{\citenamefont {Abe}\ \emph {et~al.}(2018{\natexlab{b}})\citenamefont
  {Abe} \emph {et~al.}}]{T2K:2018rnz}%
  \BibitemOpen
  \bibfield  {author} {\bibinfo {author} {\bibfnamefont {K.}~\bibnamefont
  {Abe}} \emph {et~al.} (\bibinfo {collaboration} {T2K}),\ }\href {\doibase
  10.1103/PhysRevD.98.032003} {\bibfield  {journal} {\bibinfo  {journal} {Phys.
  Rev. D}\ }\textbf {\bibinfo {volume} {98}},\ \bibinfo {pages} {032003}
  (\bibinfo {year} {2018}{\natexlab{b}})},\ \Eprint
  {http://arxiv.org/abs/1802.05078} {arXiv:1802.05078 [hep-ex]} \BibitemShut
  {NoStop}%
\bibitem [{\citenamefont {Lu}\ \emph {et~al.}(2018)\citenamefont {Lu} \emph
  {et~al.}}]{MINERvA:2018hba}%
  \BibitemOpen
  \bibfield  {author} {\bibinfo {author} {\bibfnamefont {X.~G.}\ \bibnamefont
  {Lu}} \emph {et~al.} (\bibinfo {collaboration} {MINERvA}),\ }\href {\doibase
  10.1103/PhysRevLett.121.022504} {\bibfield  {journal} {\bibinfo  {journal}
  {Phys. Rev. Lett.}\ }\textbf {\bibinfo {volume} {121}},\ \bibinfo {pages}
  {022504} (\bibinfo {year} {2018})},\ \Eprint
  {http://arxiv.org/abs/1805.05486} {arXiv:1805.05486 [hep-ex]} \BibitemShut
  {NoStop}%
\bibitem [{\citenamefont {Agostinelli}\ \emph {et~al.}(2003)\citenamefont
  {Agostinelli} \emph {et~al.}}]{AGOSTINELLI2003250}%
  \BibitemOpen
  \bibfield  {author} {\bibinfo {author} {\bibfnamefont {S.}~\bibnamefont
  {Agostinelli}} \emph {et~al.},\ }\href {\doibase
  https://doi.org/10.1016/S0168-9002(03)01368-8} {\bibfield  {journal}
  {\bibinfo  {journal} {Nuclear Instruments and Methods in Physics Research
  Section A: Accelerators, Spectrometers, Detectors and Associated Equipment}\
  }\textbf {\bibinfo {volume} {506}},\ \bibinfo {pages} {250} (\bibinfo {year}
  {2003})}\BibitemShut {NoStop}%
\bibitem [{\citenamefont {Allison}\ \emph {et~al.}(2006)\citenamefont {Allison}
  \emph {et~al.}}]{Geant4}%
  \BibitemOpen
  \bibfield  {author} {\bibinfo {author} {\bibfnamefont {J.}~\bibnamefont
  {Allison}} \emph {et~al.},\ }\href {\doibase 10.1109/TNS.2006.869826}
  {\bibfield  {journal} {\bibinfo  {journal} {IEEE Transactions on Nuclear
  Science}\ }\textbf {\bibinfo {volume} {53}},\ \bibinfo {pages} {270}
  (\bibinfo {year} {2006})}\BibitemShut {NoStop}%
\bibitem [{\citenamefont {Allison}\ \emph {et~al.}(2016)\citenamefont {Allison}
  \emph {et~al.}}]{Allison:2016lfl}%
  \BibitemOpen
  \bibfield  {author} {\bibinfo {author} {\bibfnamefont {J.}~\bibnamefont
  {Allison}} \emph {et~al.},\ }\href {\doibase 10.1016/j.nima.2016.06.125}
  {\bibfield  {journal} {\bibinfo  {journal} {Nucl. Instrum. Meth. A}\ }\textbf
  {\bibinfo {volume} {835}},\ \bibinfo {pages} {186} (\bibinfo {year}
  {2016})}\BibitemShut {NoStop}%
\end{thebibliography}%
\end{document}